\documentclass[emulateapj,apj]{emulateapj}

\usepackage{natbib}

\slugcomment{Not to appear in Nonlearned J., 45.}

\shorttitle{
Self-Sustained Turbulence in a Bistable Interstellar Medium
}
\shortauthors{Iwasaki and Inutsuka}

\begin{document}
%\tableofcontents


\title{
Self-Sustained Turbulence without Dynamical Forcing: \\
A Two-Dimensional Study of a Bistable Interstellar Medium 
}


\author{Kazunari iwasaki\altaffilmark{1} and Shu-ichiro Inutsuka\altaffilmark{1} 
}

\altaffiltext{1}{Department of Physics, Nagoya University, Furo-cho, 
Chikusa-ku, Nagoya, Aichi, 464-8602, Japan; 
iwasaki@nagoya-u.jp, inutsuka@nagoya-u.jp}


%\altaffiltext{1}{Visiting Astronomer, Cerro Tololo Inter-American Observatory.
%CTIO is operated by AURA, Inc.\ under contract to the National Science
%\altaffiltext{2}{Society of Fellows, Harvard University.}
%\altaffiltext{3}{present address: Center for Astrophysics,
%    60 Garden Street, Cambridge, MA 02138}
%\altaffiltext{4}{Visiting Programmer, Space Telescope Science Institute}
%\altaffiltext{5}{Patron, Alonso's Bar and Grill}


\begin{abstract}
In this paper, the nonlinear evolution of a bistable interstellar medium is investigated 
using two-dimensional simulations  
with a realistic cooling rate, thermal conduction, and physical viscosity.
The calculations are performed using periodic boundary conditions without any external dynamical forcing. 
As the initial condition, a spatially uniform unstable gas under thermal equilibrium is considered.
At the initial stage, the unstable gas quickly segregates into two phases, or cold neutral medium (CNM)
and warm neutral medium (WNM). 
Then, self-sustained turbulence with velocity dispersion of $0.1-0.2\;\mathrm{km\;s^{-1}}$
is observed in which the CNM moves around in the WNM. 
We find that the interfacial medium (IFM) between the CNM and WNM
plays an important role in sustaining the turbulence.
The self-sustaining mechanism can be divided into two steps.
First, thermal conduction drives fast flows streaming into concave CNM surfaces towards the WNM.
The kinetic energy of the fast flows in the IFM is incorporated into that of the CNM through the phase transition.
Second, turbulence inside the CNM deforms interfaces and forms other concave CNM surfaces, 
leading to fast flows in the IFM. This drives the first step again and 
a cycle is established by which turbulent motions are self-sustained.

\end{abstract}



%%%%%%%%%%%%%%%%%%%%%%%%%%%%%%%%%%%%%%%%%%%%%%%%%%%%%%%
\section{Introduction} \label{sec:intro}
%%%%%%%%%%%%%%%%%%%%%%%%%%%%%%%%%%%%%%%%%%%%%%%%%%%%%%%
It is well known that the interstellar medium (ISM) has
a thermally bistable structure in the optically thin regime as a result of 
the balance of radiative cooling and heating due to 
external radiation fields and cosmic rays \citep{FGH69,Wetal95,Wetal03}.
The bistable gas consists of two thermally equilibrium phases, i.e.,
a clumpy low-temperature phase [cold neutral medium (CNM)] and 
a diffuse high-temperature phase [warm neutral medium (WNM)].
The CNM is observed as HI clouds ($n\sim10-100\;\mathrm{cm^{-3}}$, $T\sim10^2$ K), 
and the WNM is observed as diffuse HI gas ($n\sim0.1\;\mathrm{cm^{-3}}$, $T\sim 6000$ K).
In the temperature range between these phases, the gas is thermally unstable.

Linear analyses of the thermal instability (TI) have been investigated by \citet{F65} for 
a uniform gas under thermal equilibrium and by \citet{B86} for thermal nonequilibrium gas.
They found criteria for the TI.
\citet{IT08} discovered a one parameter family of self-similar solutions which describe 
the nonlinear development of the TI for various scales under a plane-parallel geometry.
Their linear stability was investigated by \citet{IT09}.

The basic physics of bistable gas has been investigated by many authors. \citet{ZP69}
investigated the steady state structure of a transition layer connecting the CNM and WNM 
under a plane-parallel geometry \citep[see also][for a larger parameter space]{II12}.
The thickness of the transition layer corresponds to the Field length,
below which 
the TI is stabilized by thermal conduction \citep{F65}. 
They found a so-called saturation 
pressure $P_\mathrm{sat}$ at which there is a static solution.
If the surrounding pressure is larger (smaller) than $P_\mathrm{sat}$, the solution describes 
condensation (CNM$\leftarrow$WNM)  (evaporation (CNM$\rightarrow$WNM)).
\citet{YT09} discovered pulselike static solutions and demonstrated that they are sustained by the 
balance between viscosity and the pressure gradient.
\citet{ERS91,ERS92} have investigated the interaction between multi-transition layers.
They found that the transition layers tend to approach and annihilate.
The merging timescale is an exponentially increasing function of the separation between the transition layers.
\citet{AMS93} investigated the nonlinear evolution of 
a thermally unstable phase under a plane-parallel 
geometry for the case of open boundaries. 
In the early phase, runaway condensation occurs in 
dense regions and rarefied parts are heated up until both reach thermal equilibria. 
Finally, the pressure approaches the saturation pressure \citep{ZP69}.
Linear analysis of a plane-parallel transition layer has been done by \citet{AMS95} 
for the long-wavelength limit including curvature effects and
by \citet{IIK06} for the long- and short-wavelength limits. They found that an evaporation front is unstable 
against corrugation-type fluctuations while a condensation front is stable.
Recently, \citet{KK13} have investigated the nonlinear development of the evaporation-front instability.
\citet{SZ09} have shown the presence of magnetic fields perpendicular to transition layers modifies 
their stability properties.

The multi-dimensional dynamics of a bistable gas is quite different 
from the one-dimensional case.
\citet{GL73} found a minimum cloud size below which clouds inevitably evaporate by 
investigating isobaric flows in a spherical symmetrical geometry.
\citet{ERS91} found that a transition layer at $P=P_\mathrm{sat}$
is not static in the multi-dimensional case. 
\citet{NKI05,NIK06} investigated the evaporation and condensation of a spherical and cylindrical CNM surrounded 
by a WNM under the isobaric approximation. 
The front velocity is proportional to the inverse of the radius 
at $P=P_\mathrm{sat}$,
and is constant in the case of much larger clouds and/or pressure far from $P_\mathrm{sat}$.
Those results indicate that the motion of a transition layer depends on its curvature, suggesting 
that the multi-dimensional structure is more complex than the 1D structure.

Numerical hydrodynamical simulations are powerful tools to investigate 
the multi-dimensional 
evolution of bistable gas since analytic analyses are quite difficult in
general situations.
\citet{KI06} have investigated the nonlinear evolution of bistable gas by using 
two- and three-dimensional numerical simulations incorporating a realistic cooling rate 
with periodic boundary conditions. 
They used realistic thermal conduction and viscosity in their fiducial model.
Interestingly, they found self-sustained turbulence in bistable gas even though
they did not consider any external dynamical forcing.
On the other hand, \citet{Brandenburg07} have performed 
similar calculations and concluded that there is no sustained turbulence.
However, to resolve the thickness of the interface, 
they adopted an artificially large thermal conductivity so that 
the Field length was as large as $\sim 0.5$pc and spatially constant.
The actual Field length has a large spatial variation.
It is as small as $\sim\;\mathrm{several}\times10^{-3}\;\mathrm{pc}$ 
for the CNM while it is as large as $\sim 0.1$pc for the WNM.
Moreover, they considered an artificially large viscosity whose value 
is determined so that the Prandtl number is unity. 
Because of such overly large viscosity, turbulence may decay in their simulation.

To understand whether turbulence is sustained or not, 
it is important to understand its driving mechanism.
Energetically it is possible because there is a continuous energy input by external heating.
However, the detailed mechanism is still unknown \citep{KI06}.
In this paper, the detailed turbulent structure of bistable
gas is investigated in order to understand the driving mechanism of turbulence. 
To obtain converged results, the thickness of the
transition layer needs to be resolved by at least a few grids \citep{KI04,KK13}. Since the required grid size is less than 
$\sim\:\mathrm{several}\times10^{-3}\;\mathrm{pc}$, it is computationally quite expensive 
to perform three-dimensional simulations even if the simulation box is as small as several pc.
Thus, as a first step, the two-dimensional evolution is considered with sufficient resolution.

This paper is organized as follows. The basic equations and numerical methods are described in 
Section \ref{sec:basic}.
The results of the two-dimensional simulations are shown in Section \ref{sec:result}.
Our results are discussed in Section \ref{sec:discuss} and summarized in Section \ref{sec:summary}.

%%%%%%%%%%%%%%%%%%%%%%%%%%%%%%%%%%%%%%%%%%%%%%%%%%%%%%%
\section{\label{sec:basic}Equations and Methods} 
%%%%%%%%%%%%%%%%%%%%%%%%%%%%%%%%%%%%%%%%%%%%%%%%%%%%%%%
%------------------------------
\subsection{Basic Equations}
%------------------------------
The Navier-Stokes equations with radiative cooling/heating and thermal conduction are solved, 
\begin{equation}
        \frac{\partial\rho}{\partial t} + \frac{\partial}{\partial x_j} \left( \rho v_j \right)=0,
        \label{eoc}
\end{equation}
\begin{equation}
  \frac{\partial \left(\rho v_j\right)}{\partial t} + \frac{\partial}{\partial x_j} 
  \left( P \delta_{ij} + \rho v_i v_j - \sigma_{ij} \right)=0,
  \label{eom}
\end{equation}
\begin{equation}
        \frac{\partial E}{\partial t} + \frac{\partial}{\partial x_j} 
        \left[ \left( E+P \right)v_j - \sigma_{ij} v_i - \kappa \frac{\partial T}{\partial x_j} \right] = 
        -\rho {\cal L}(\rho,T),
        \label{eoe}
\end{equation}
\begin{equation}
        P = \frac{k_\mathrm{B}}{\mu_\mathrm{H} m_\mathrm{H}} \rho T=n_\mathrm{H} k_\mathrm{B}T,
        \label{eos}
\end{equation}
\begin{equation}
        \sigma_{ij} = \mu
        \left[ \left( \frac{\partial v_i}{\partial x_j} + \frac{\partial v_j}{\partial x_i}
        \right) - \frac{2}{3}\delta_{ij} \frac{\partial v_k}{\partial x_k} \right],
        \label{viscous}
\end{equation}
where $E=P/(\gamma-1) + \rho \vec{v}^2/2$ is the total energy, $\gamma=5/3$ is the ratio of specific heats,
$n_\mathrm{H}$ is the number density of Hydrogen nuclei, 
$\mu_{\mathrm{H}}=1.4$ is the mean molecular weight per Hydrogen nucleus,
$\mu$ is the viscosity coefficient, 
$\kappa$ is the thermal conductivity, and $\cal L$ is a net cooling rate per unit mass.
In this paper, the following simple analytic formula \citep{KI02} is adopted:
\begin{equation}
        \rho {\cal L}(\rho,T)=n_\mathrm{H}\left\{n_\mathrm{H}\Lambda(T)-\Gamma \right\}\;\;\mathrm{erg\;cm^{-3}\;s^{-1}}
\end{equation}
\[
        \Gamma = 2\times10^{-26}\;\;\mathrm{erg\;s^{-1}}
        \]
        \[
        \frac{\Lambda(T)}{\Gamma}=10^7\exp\left[ -\frac{118400}{T+1000} \right]
        + 1.4\times10^{-2}\sqrt{T}\exp\left( -\frac{92}{T} \right).
        \]
The validity of the cooling function in multi-dimensional simulation is analyzed
in a more detailed treatment \citep{Metal13}.
The thermal conductivity for neutral hydrogen 
($\kappa(T)=2.5\times10^3\sqrt{T}\;\mathrm{cm^{-1}\; K^{-1}\;s^{-1}}$) is adopted.
In a neutral monatomic gas, the viscosity coefficient is given by 
$\mu = 3\kappa/(2c_p)$, where $c_p = \gamma k_\mathrm{B}/
\{\mu_\mathrm{H}m_\mathrm{H}(\gamma-1)\}$ is the specific heat at constant pressure.

%The relation between the viscosity coefficient, $\mu$,

%======================================================================
\subsection{Thermal Properties of the ISM}\label{sec:property}
%======================================================================
\begin{figure}[htpb]
        \begin{center}
                \includegraphics[width=7cm]{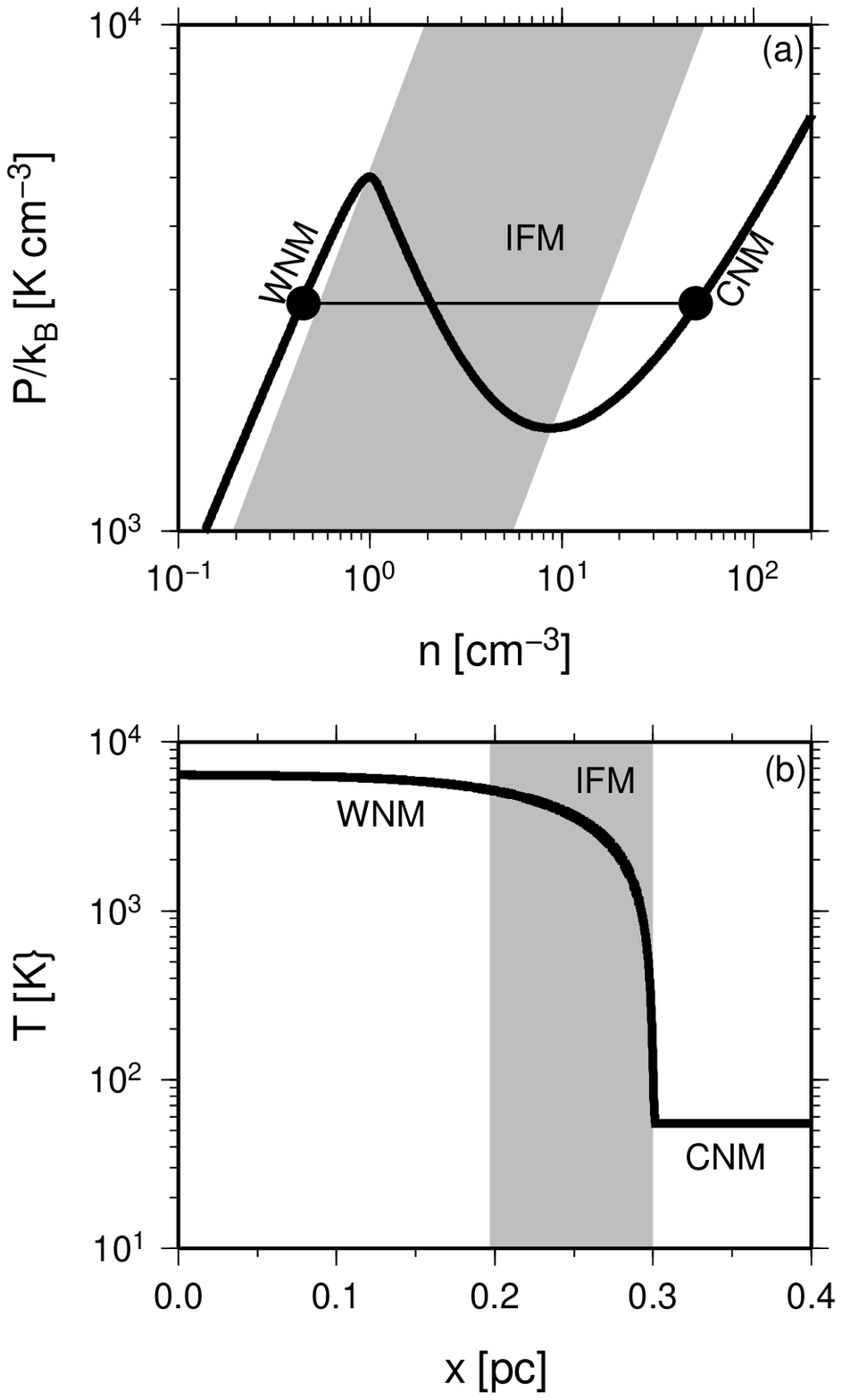}
        \end{center}
        \caption{
        {\it Upper panel}: Thermal equilibrium state in the $(n,P)$ plane {\it (solid line)}.
        {\it Lower panel}: Temperature distribution of a steady solution.
        In each panel, the gray region indicates the IFM where $(\partial {\cal L}/\partial \rho)_P>0$.
        }
        \label{fig:eq}
\end{figure}

Fig. \ref{fig:eq}a shows the thermal equilibrium curve where ${\cal L}=0$ in the $(n,P)$ plane.
One can see that the fluid can take two stable equilibrium states, 
the CNM and WNM, at constant pressure, as shown by the thin horizontal line.
In a bistable fluid, an interfacial medium (IFM) connects the CNM and 
WNM. In this paper, the IFM is defined as the gas in the gray region in Fig. \ref{fig:eq}a where 
$\left(\partial {\cal L}/\partial \rho\right)_P > 0$.
\citet{ZP69} found steady solutions connecting the CNM/WNM. 
Fig. \ref{fig:eq}b shows the temperature distribution of a steady solution.
The gray region indicates the IFM corresponding to 
a transition layer as discussed in Section \ref{sec:intro}.
The thickness of the IFM is characterized by the Field length 
\citep{BM90},
\begin{equation}
        \lambda_\mathrm{F} = \sqrt{\frac{\kappa(T)T}{\rho |{\cal L}|}}.
        \label{}
\end{equation}
The Field length depends on local temperatures and densities. The Field length for the CNM is as small 
as several $\times10^{-3}\;\mathrm{pc}$, while for the WNM it is as large as 
$0.1\;\mathrm{pc}$.
This dependence can be seen in Fig. \ref{fig:eq}b.
From the WNM, the temperature gradually declines because of the large Field length.
As the temperature decreases, the Field length decreases so the temperature rapidly 
drops to CNM values.
Thus, the CNM and IFM are separated by a sharp discontinuity that hereafter we will refer to as an interface 
or CNM surface.
On the other hand, 
the WNM is smoothly connected with the IFM, as shown in Fig. \ref{fig:eq}b.

%======================================================================
\subsection{Methods and Initial Conditions}
%======================================================================
An operator-splitting technique is used for solving the basic equations (\ref{eoc})-(\ref{eoe}).
For the inviscid part, a second-order Eulerian-remap Godunov scheme \citep{vL79} is used.
The cooling/heating, thermal conduction, and physical viscosity 
are calculated by explicit time integration. 
A square domain $-L/2<x,\;y<L/2$ is considered, where $L$ is the domain length.
Periodic boundary conditions are imposed in the $x$- and $y$-directions.

As an initial condition, a uniform unstable gas 
($n_\mathrm{H}=4.3$ cm$^{-3}$ and $T=423$ K)
in thermal equilibrium is considered. The initial state is in the IFM phase.
A random velocity fluctuation with
a flat power spectrum whose minimum scale is $L/4$ is added to the initial state.
The amplitude of the velocity dispersion is $\sim$2\% of the sound speed.
It has been confirmed that saturation levels of turbulence do not depend on how 
initial fluctuations are added. 
As a fiducial model, a case with $L=2.4\;\mathrm{pc}$ is considered.
The box size dependence of turbulence will be investigated in Section \ref{sec:Ldep}.
\citet{KI04} have proposed the Field condition where the local Field length should be resolved by 
a few grids to obtain the converged results \citep[see also][]{KK13}. 
Thus, the minimum Field length of $\sim 3\times 10^{-3}\;\mathrm{pc}$ in the CNM needs to be resolved.
In the fiducial model, $N=2048^2$ is used, where $N$ is the total cell number. The corresponding grid size 
is $10^{-3}\;\mathrm{pc}$ that satisfies the Field condition.

%======================================================================
\section{Results}\label{sec:result}
%======================================================================
%-----------------------------------
\subsection{Velocity Dispersion}\label{sec:veldisp}
%-----------------------------------
\begin{figure}[htpb]
        \begin{center}
            \includegraphics[width=7cm]{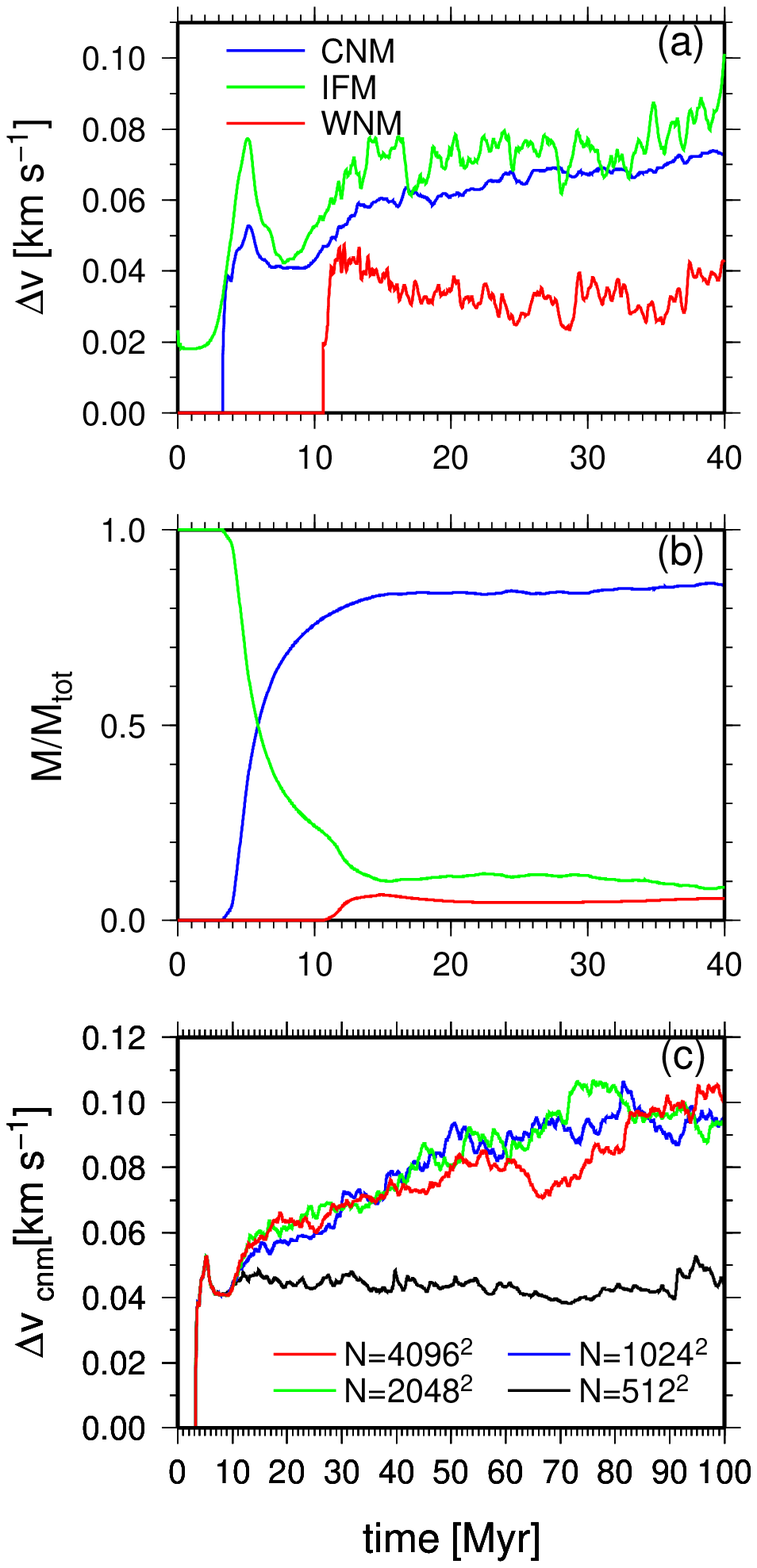}
        \end{center}
        \caption{Early evolution for $t<20\;\mathrm{Myr}$ of (a)velocity dispersion and 
        (b)mass fraction for the three phases, 
        the CNM {\it (blue)}, IFM {\it (green)}, and WNM {\it(red)}.
        (c)resolution dependence of $\Delta v_\mathrm{cnm}$.
        The red, green, blue, and black lines correspond to the results with 
        $N=4096^2$, $2048^2$, $1024^2$, and $N=512^2$, respectively.
        }
        \label{fig:vel_ini}
\end{figure}

\begin{figure}[htpb]
        \begin{center}
            \includegraphics[width=9cm]{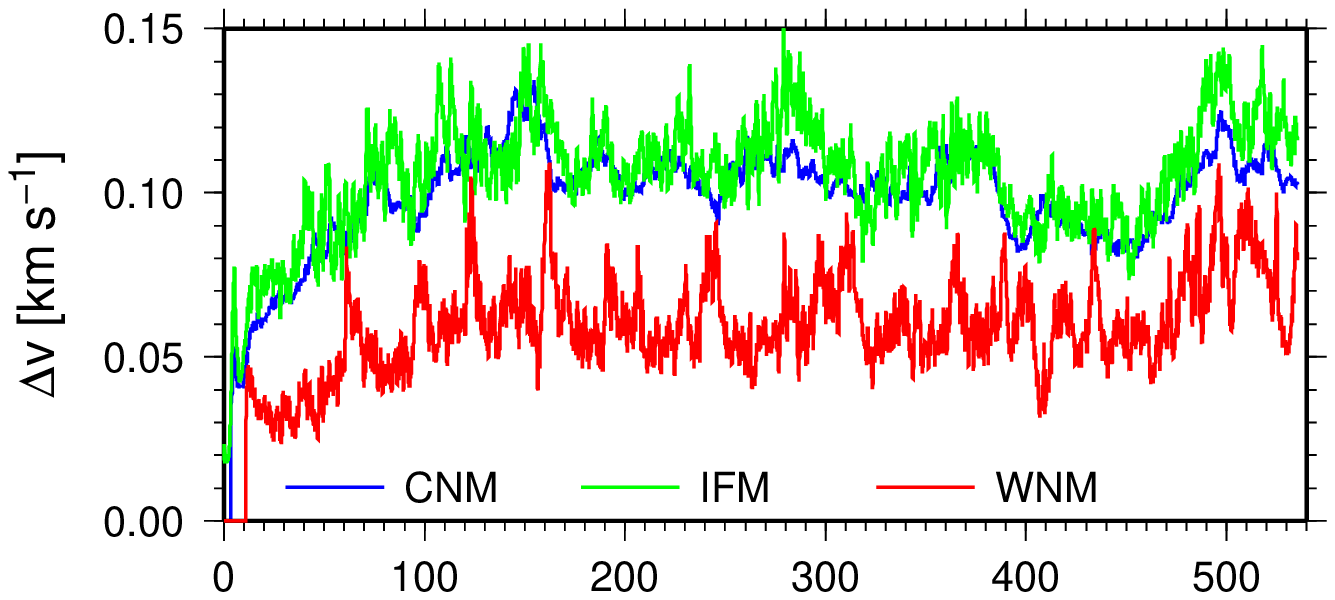}
        \end{center}
        \caption{The same as Fig. \ref{fig:vel_ini}a but for the long-term evolution.
        }
        \label{fig:vel_las}
\end{figure}

In this section, the time evolution of the velocity dispersions and 
mass fractions of the three phases (CNM, IFM, WNM) are investigated.
The phase of each grid cell is distinguished using Fig. \ref{fig:eq}a.
The mass and velocity dispersion of each phase are given by 
\begin{equation}
        M_\mathrm{s}=\int_{V_\mathrm{s}} \rho dV,
\end{equation}
and
\begin{equation}
    \Delta v_\mathrm{s} \equiv \sqrt{ \frac{1}{M_\mathrm{s}} \int_\mathrm{V_\mathrm{s}}\rho {\bf v}^2 dV},
\end{equation}
respectively, where the subscript ``s'' denotes the phase, $s=\mathrm{(cnm,\;ifm,\;wnm)}$,
and $V_\mathrm{s}$ indicates 
the volume occupied by the phase ``s''.

Fig. \ref{fig:vel_ini}a and \ref{fig:vel_ini}b show 
the early evolution for $t<40\;\mathrm{Myr}$ of the velocity dispersion and 
mass fraction, $M_\mathrm{s}/M_\mathrm{tot}$, of the three phases, respectively, where 
$M_\mathrm{tot}$ is the total mass.
Initially, the TI causes runaway cooling in the dense regions while
runaway heating in the rarefied parts keeps the pressure almost constant.
During this time, the velocity dispersion of the IFM increases 
exponentially (see Fig. \ref{fig:vel_ini}a).
From Fig. \ref{fig:vel_ini}b, one can see that $M_\mathrm{ifm}$ begins
to decrease while $M_\mathrm{cnm}$ quickly increases around $t\sim 4$ Myr. 
This indicates that the dense parts of the IFM change into the CNM.
Around $t\sim 15$ Myr, $M_\mathrm{cnm}/M_\mathrm{tot}$ reaches $\sim 80\%$.
The formation epoch of the WNM lags behind that of the CNM because the heating timescale in the rarefied parts
is longer than the cooling timescale in the dense parts.
Around $t\sim 20$ Myr, a bistable fluid consisting of the CNM/WNM
is formed. The IFM occupies the regions inbetween. The mass fraction 
of the three phases is 
$M_\mathrm{cnm}:M_\mathrm{ifm}:M_\mathrm{wnm}\sim 0.85:0.10:0.05$.

Before proceeding, the resolution dependence of 
the velocity dispersion is investigated.
Fig. \ref{fig:vel_ini}c shows the velocity dispersion of CNM for $N=512^2(\Delta x=1.6\lambda_\mathrm{F,min})$, 
$1024^2(\Delta x=0.8\lambda_\mathrm{F,min})$, $2048^2(\Delta x=0.4\lambda_\mathrm{F,min})$,
$4096^2(\Delta x=0.2\lambda_\mathrm{F,min})$, where $\lambda_\mathrm{F,min}=3\times10^{-3}\;\mathrm{pc}$ is 
the minimum Field length.
From Fig. \ref{fig:vel_ini}c, for $t<10\;\mathrm{Myr}$ when the TI develops, the velocity dispersion is independent of 
resolution. This is because all models resolve the maximum growth scale of the TI $(\sim0.04\;\mathrm{pc})$.
After the bistable gas is formed, the WNM/CNM are separated by 
the transition layers whose thicknesses near the CNM correspond to $\lambda_\mathrm{F,min}$ (see Section \ref{sec:property}).
For the lowest resolution case ($N=512^2$), $\lambda_\mathrm{F,min}$ is not resolved.
That is why only the result with $N=512^2$ exhibits the lowest velocity dispersion and turbulence does not increases with time.
On the other hand, $\Delta v_\mathrm{cnm}$
for the three higher resolution models ($N=1024^2,2048^2$, and $4096^2$) reach 
almost the same values at $t=100\;\mathrm{Myr}$ and the results appear to be converged.
The two models ($N=2048^2$ and $4096^2$) satisfy the Field condition while the model with $N=1024^2$ barely 
resolves $\lambda_\mathrm{F,min}$. These results are consistent with \citet{KI04}.
Thus, the fiducial model produces the converged result.

Next, the long term evolution of the bistable fluid is investigated.
Fig. \ref{fig:vel_las} is the same as Fig. \ref{fig:vel_ini}a but for 
$40<t/\mathrm{Myr}<530$.
It is found that turbulence is maintained until at least $530$ Myr in all phases.
The evolution of $\Delta v_\mathrm{cnm}$ is quite similar to 
that of $\Delta v_\mathrm{ifm}$, while $\Delta v_\mathrm{ifm}$ has larger fluctuations.
The time-averaged velocity dispersion is as large as $\sim 0.1\;\mathrm{km\;s^{-1}}$.
On the other hand, $\Delta v_\mathrm{wnm}$ is smaller than the other two phases.
Most of the kinetic energy of the turbulence resides in the CNM 
because of its large mass fraction. 
If there is only CNM, the turbulence is expected to decay within its crossing timescale 
$\sim L/(0.1\;\mathrm{km\;s^{-1}}) \sim 23\;\mathrm{Myr}$. 
Thus, maintenance of the turbulence requires a supply of kinetic energy 
into the CNM. 

%------------------------------------------------
\subsection{Density and Velocity Distributions}\label{sec:denvel}
%------------------------------------------------
\begin{figure*}[htpb]
        \begin{center}
                \includegraphics[width=16cm]{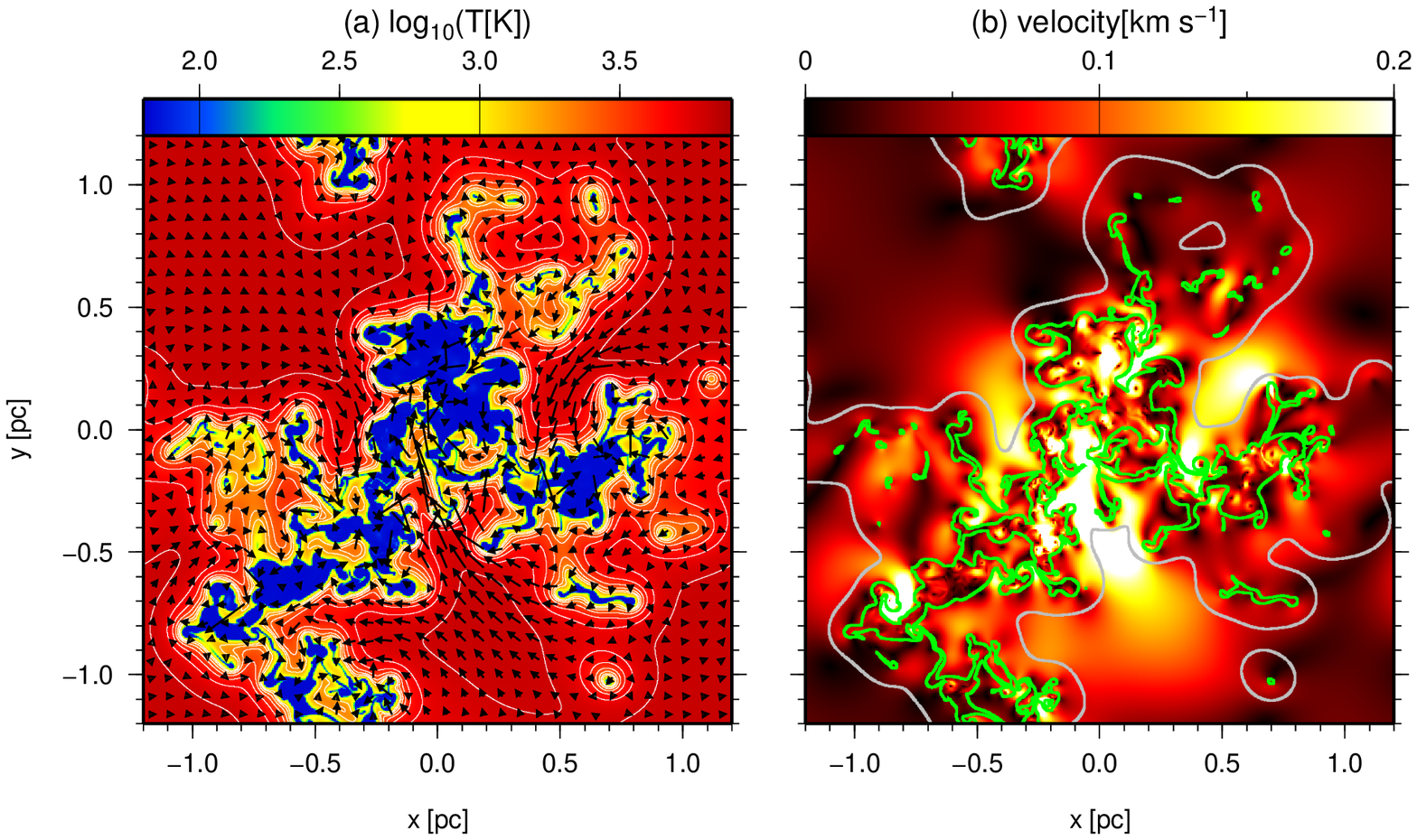}
        \end{center}
        \caption{
        (a)Color and contour maps of temperature at a fixed epoch. 
        The contour interval is 0.1 with values between 1.8 and 3.8 in $\log_{10}(T[K])$.
        The arrows show the velocity field.
        (b)Color map of the velocity amplitude. The green (gray) line corresponds to 
        the boundary between CNM/IFM (WNM/IFM). The regions between these two lines correspond 
        to the IFM.
        }
        \label{fig:denvel}
\end{figure*}

\begin{figure}[htpb]
        \begin{center}
                \includegraphics[width=8cm]{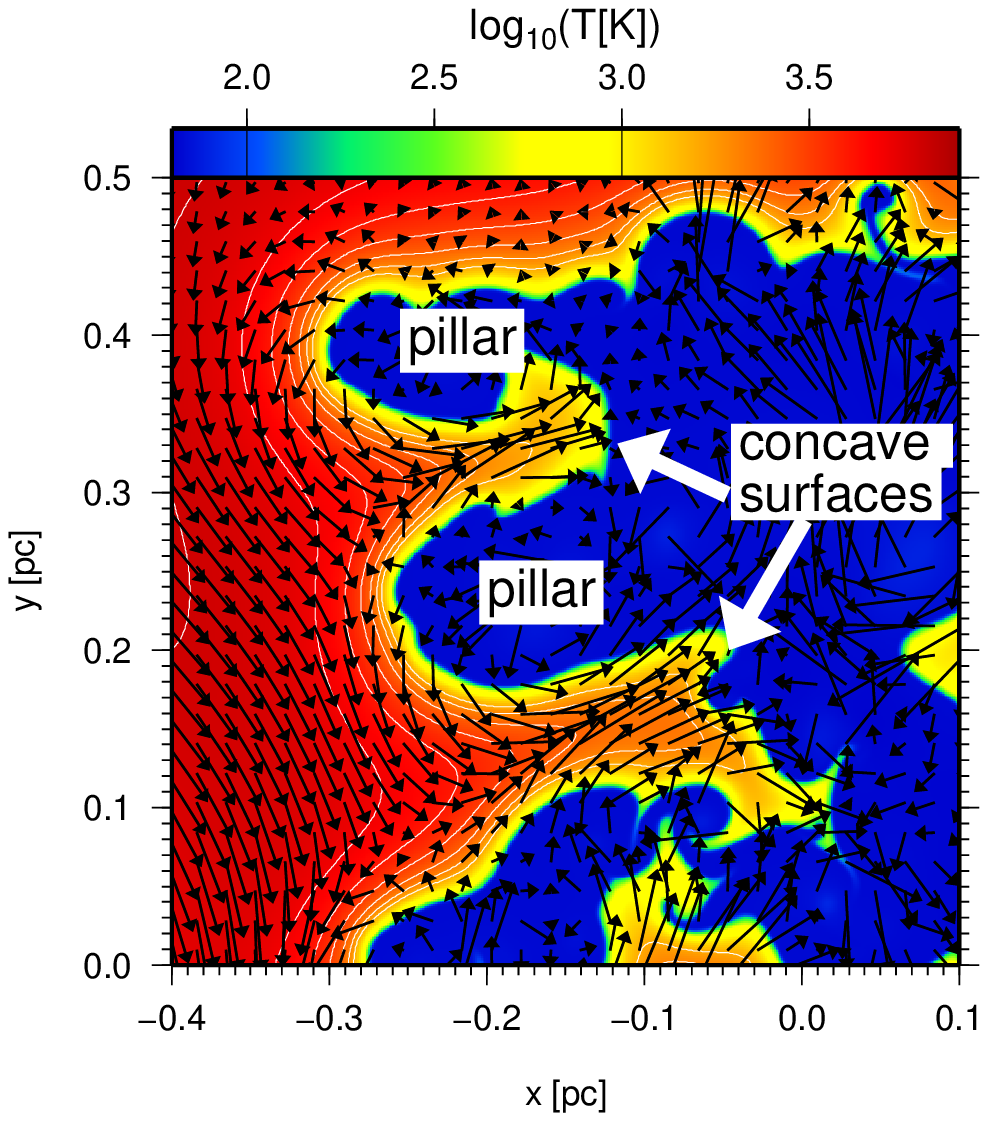}
        \end{center}
        \caption{
        Close-up view of Fig. \ref{fig:denvel}a.
        }
        \label{fig:denzoom}
\end{figure}

Fig. \ref{fig:denvel}a shows color and contour maps of the temperature at a fixed epoch.
One can see that the CNM (blue) has a complicated structure in the 
WNM (red). To see the turbulent structure, the 
color map of the velocity amplitude $|\bf v|$ is shown in Fig. \ref{fig:denvel}b.
The green lines correspond to CNM/IFM interfaces and 
the gray lines indicate IFM/WNM boundaries.
Thus, the regions between these lines belong to the IFM.
Fig. \ref{fig:denvel}b shows that the CNM has a complicated fine velocity structure 
while the WNM does not.
This comes from the large difference of the Reynolds numbers, $L\Delta v/\nu$, of the WNM and CNM that 
are given by 
\begin{eqnarray}
   \mathrm{Re_{wnm}} &=& 60\left( \frac{n_\mathrm{wnm}}{0.5\;\mathrm{cm}^{-3}} \right) 
   \left( \frac{T_\mathrm{wnm}}{6000\;\mathrm{K}} \right)^{-1/2} \nonumber \\
  && 
 \hspace{1.8cm}\times\left( \frac{\Delta v_\mathrm{wnm}}{0.06\;\mathrm{km\;s^{-1}}} \right) \left( \frac{L}{2.4\;\mathrm{pc}} \right),
\end{eqnarray}
and
\begin{eqnarray}
    \mathrm{Re_{cnm}} &=& 10^5\left( \frac{n_\mathrm{cnm}}{50\;\mathrm{cm}^{-3}} \right) 
  \left( \frac{T_\mathrm{cnm}}{50\;\mathrm{K}} \right)^{-1/2} \nonumber\\
  &&\hspace{2cm}\times\left( \frac{\Delta v_\mathrm{cnm}}{0.1\;\mathrm{km\;s^{-1}}} \right) \left( \frac{L}{2.4\;\mathrm{pc}} \right),
\end{eqnarray}
respectively, where $\Delta v_\mathrm{wnm}$ and $\Delta v_\mathrm{cnm}$ are evaluated in Fig. \ref{fig:vel_las}.
%The corresponding Kolmogorov scales of the WNM and CNM are $0.04\;\mathrm{pc}$ and $2.4\times10^{-5}\;\mathrm{pc}$, respectively.
One can see that the Reynolds number of WNM is much smaller than that of CNM by about three oder of magnitude.
Thus, the turbulent CNM is embedded in the viscous WNM.
This dissipative feature of the WNM is also seen in Fig. \ref{fig:vel_las} where 
the WNM has the smallest velocity dispersion.
Note that fast flows are seen in the IFM near the deformed CNM/IFM interfaces.

Fig. \ref{fig:denzoom} shows a close-up view of Fig. \ref{fig:denvel}a.
There are two prominent types of strongly curved CNM surfaces.
One is a deep concave CNM surface towards the WNM.
The other is a pillar corresponding to an elongated convex surface towards the WNM.
From Fig. \ref{fig:denzoom}, it is seen that the fast flows in the IFM flood into the 
concave CNM surface while the gases in the IFM stream into the WNM from the heads of the pillars.

%%-------------------------------------------------------------------------
\subsection{Driving Mechanism of Fast Flows in the IFM}\label{sec:selfsus}
%%-------------------------------------------------------------------------
%%%%
\begin{figure*}[htpb]
        \begin{center}
                \includegraphics[width=16cm]{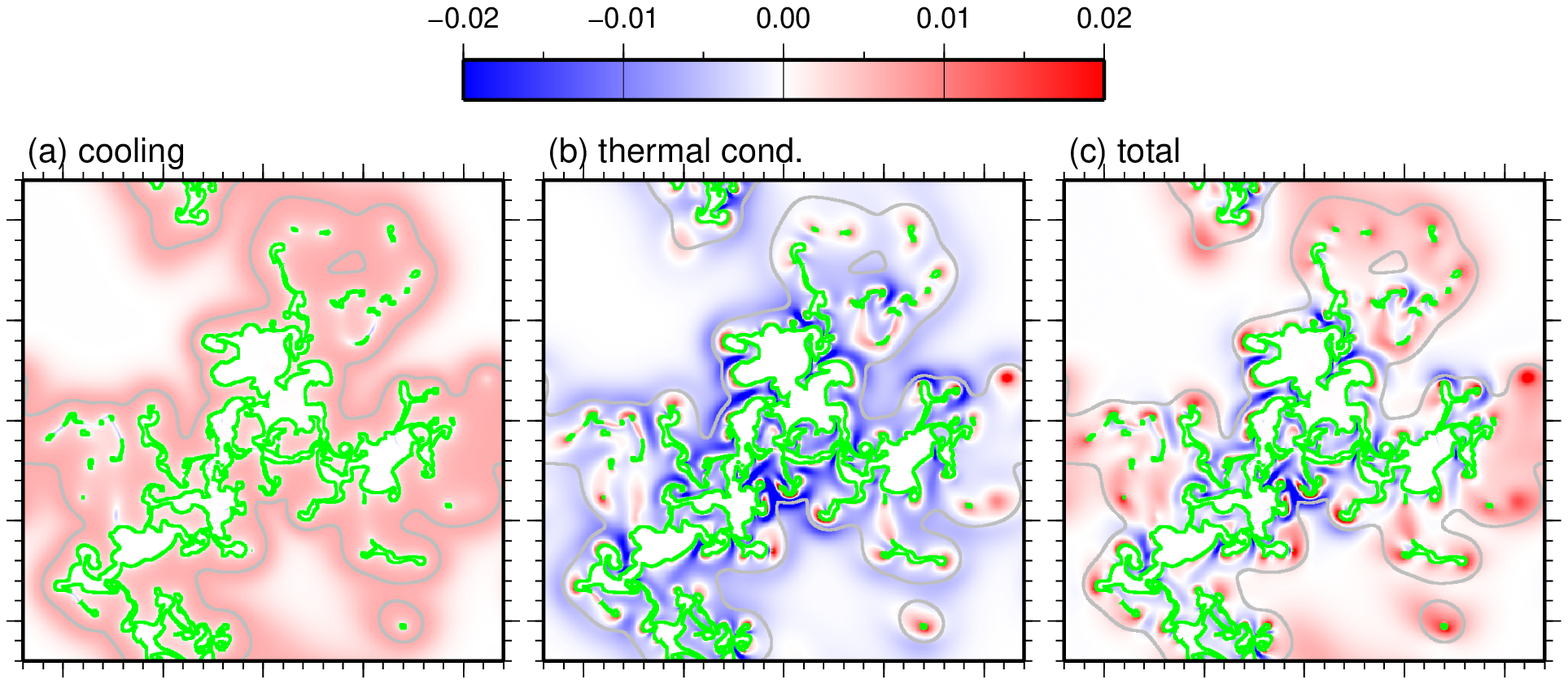}
        \end{center}
        \caption{
        Color maps of (a)net cooling rate, $-{\cal L}$, 
        (b)thermal conduction, ${\bf \nabla}\cdot \left( \kappa{\bf \nabla}T \right)/\rho$, and 
        (c)the sum of the net cooling rate and thermal conduction. 
        }
        \label{fig:ent}
\end{figure*}
\begin{figure*}[htpb]
        \begin{center}
                \includegraphics[width=11.0cm]{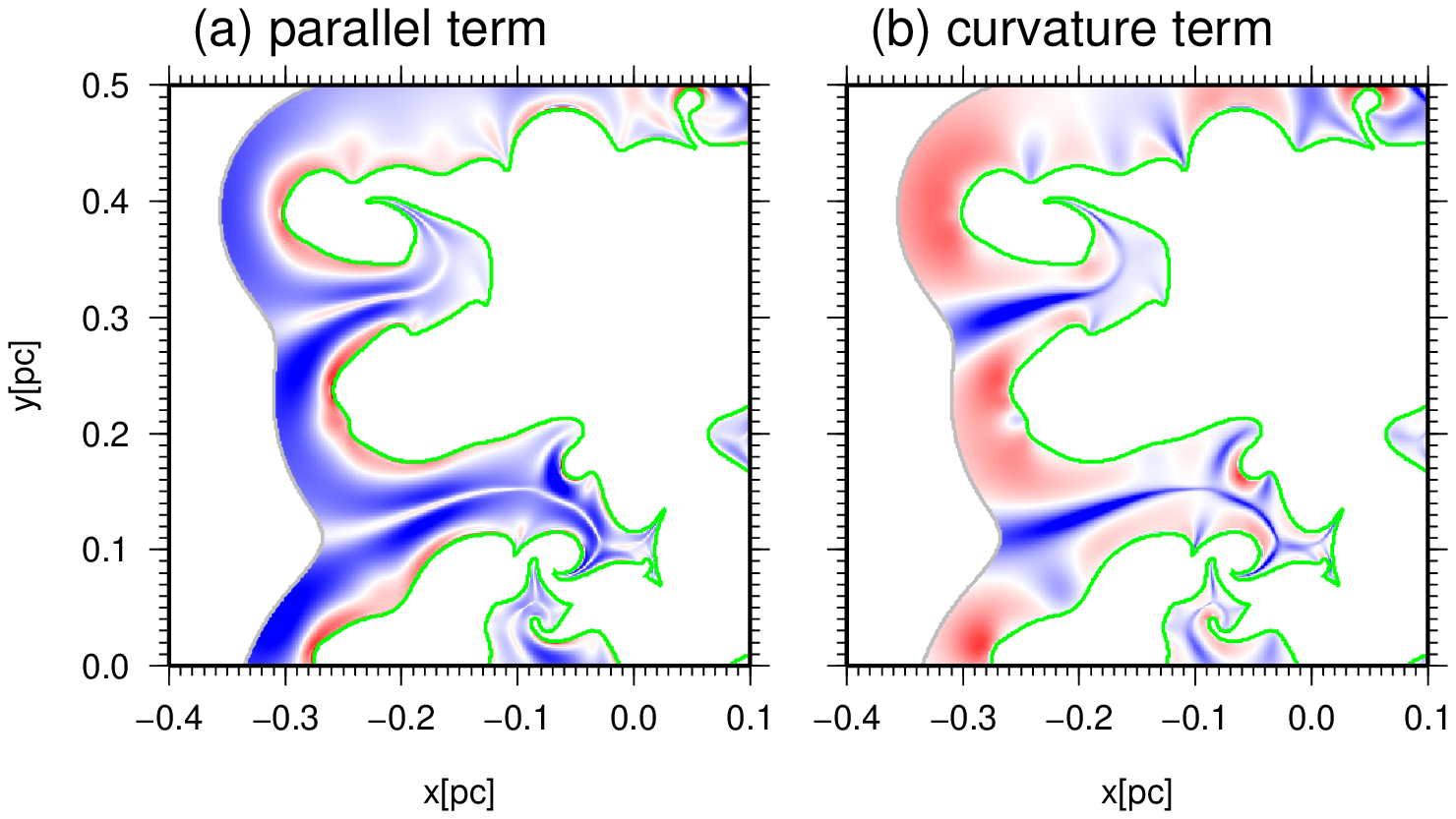}
        \end{center}
        \caption{
        Color maps of (a)the parallel term, $\partial_n\left( \kappa \partial_n T \right)$, and 
        (b)the curvature term, $\kappa K \partial_n T$, of thermal conduction in the same region 
        as Fig. \ref{fig:denzoom}.
        The green and grey lines and color bar are the same as in Fig. \ref{fig:ent}.
        }
        \label{fig:curv}
\end{figure*}
%%%%

It is well known that in a bistable gas, flows can be driven by thermal processes, i.e.,
thermal conduction and radiative cooling/heating.
This relates to the phase transition \citep{ZP69}.
The time evolution of the enthalpy is given by 
\begin{equation}
        \frac{d}{dt}\left( c_p T\right)
        = -\frac{1}{\rho} \frac{dP}{dt} + 
        \frac{1}{\rho}{\bf \nabla}\cdot\left(\kappa {\bf \nabla}T  \right) - {\cal L},
        \label{ent}
\end{equation}
where the viscous heating term is implicitly neglected because it is much smaller than the other terms.
Furthermore, the first term on the right-hand side of equation (\ref{ent}) is negligible compared with 
the left-hand side term.

Figs. \ref{fig:ent}a and \ref{fig:ent}b show color maps of the net cooling term, $-{\cal L}$, and 
the thermal conduction term, ${\bf \nabla}\cdot\left( \kappa {\bf \nabla}T \right)/\rho$,
respectively. 
Fig. \ref{fig:ent}c shows the sum of the two terms, corresponding to $d(c_p T)/dt$.
In each panel, the gas in the red (blue) region is heated (cooled).
Fig. \ref{fig:ent}a shows that most of the volume of the IFM is heated by the external radiation.
Although the gases cool in very thin layers just outside the CNM, they are too thin to be seen 
in the figure.
On the other hand, the thermal conduction term has a complicated distribution, as shown in Fig. \ref{fig:ent}b.
The thermal conduction term is positive (negative) in the IFM near the convex (concave) CNM surfaces.
Fig. \ref{fig:ent}c shows that the distribution of thermal conduction is preserved in that of 
$d(c_pT)/dt$, indicating that thermal conduction dominates the thermal process.

The distribution of thermal conduction reflects the complicated temperature distribution 
(see Fig. \ref{fig:denvel}a).
We now consider an individual temperature contour.
The unit vector parallel to the gradient vector ${\bf \nabla}T$ at a point 
on the contour is defined as ${\bf n} = {\bf \nabla}T/|{\bf \nabla} T|$. 
The vector ${\bf n}$ is oriented in the direction from the CNM to the WNM.
Using ${\bf n}$, one can write ${\bf \nabla}T = \left(\partial_n T\right) {\bf n}$, where 
$\partial_n \equiv {\bf n}\cdot{\bf\nabla}$.
The thermal conduction term can be rewritten as
\begin{equation}
  {\bf \nabla} \cdot \left(\kappa {\bf \nabla}T\right) = 
  \partial_n \left( \kappa \partial_n T \right) + \kappa K \partial_n T,
  \label{curv}
\end{equation}
where $K\equiv {\bf \nabla}\cdot{\bf n}$ is the curvature of the contour line.
The first term on the right-hand side of equation (\ref{curv}) corresponds to the 
contribution from the component parallel to {\bf n}.
The second term comes from the curvature effect which corresponds to 
the spatial variation of {\bf n}. Figs. \ref{fig:curv}a and \ref{fig:curv}b 
show the parallel term ($\partial_n \left( \kappa \partial_n T \right)/\rho$) and 
curvature term ($\kappa K \partial_n T/\rho$) in the IFM, respectively. 
From Fig. \ref{fig:curv}a, in most parts of the IFM the parallel term is negative.
Although this term is positive in the very thin layers just outside the CNM, they are too thin to be seen 
in Fig. \ref{fig:curv}a. 
In the IFM near strongly curved interfaces, $|\partial_n \left( \kappa \partial_n T \right)|$ 
is significantly enhanced because the deformation of the interfaces modifies 
the temperature distribution in the IFM.
The parallel term is small in narrow valleys in the temperature 
distribution near the concave CNM surface, as shown in Fig. \ref{fig:denvel}a.

Next, the curvature term in Fig. \ref{fig:curv}b is considered.
Its sign is determined by that of $K$ since $\partial_n T$ is positive.
At a convex (concave) contour of the temperature, the curvature term is positive (negative) as shown in Figs.
\ref{fig:denvel}a and \ref{fig:curv}b.
By comparing Figs. \ref{fig:curv}a and \ref{fig:curv}b, one can see that the curvature term is dominated near 
the pillars where the thermal conduction term is positive.
In regions with negative thermal conduction, the curvature term is important in the narrow valleys 
where the parallel term is small. The parallel term is important on both sides of the valleys. 

The typical IFM flow velocity driven by thermal conduction is estimated.
If a quasi-steady state is assumed, equation (\ref{ent}) becomes
%%%%
\begin{equation}
        c_p (\partial_n T)v_n\simeq 
        \frac{1}{\rho} K \kappa \partial_n T,
        \label{ent1}
\end{equation}
%%%%
where $v_n$ is the gas velocity parallel to ${\bf n}$, the net cooling rate is neglected, and
only regions where the curvature term is dominated are considered.  
From equation (\ref{ent1}), the typical velocity $v_n$ is estimated by 
%%%%
\begin{eqnarray}
        |v_\mathrm{n}|\sim \frac{\kappa(T_\mathrm{ifm}) |K|}{c_p \rho_\mathrm{ifm}}
        =0.2\;\mathrm{km\;s^{-1}} && \nonumber\\
        && \hspace{-3.8cm}\times\left( \frac{T_\mathrm{ifm}}{2000\;\mathrm{K}} \right)^{1/2}
        \left( \frac{|K|}{200\;\mathrm{pc^{-1}}} \right)
        \left( \frac{n_\mathrm{ifm}}{1\; \mathrm{cm^{-3}}} \right)^{-1},
   \label{curve}
\end{eqnarray}
where physical values typical of the IFM are used and a typical 
curvature value is derived from Fig. \ref{fig:curv}b.
This typical velocity, $v_\mathrm{n}$, is consistent with that found in Fig. \ref{fig:denvel}b.
%The flow direction is determined by the sign of $K$. 
%The regions where $K>0$ ($K<0$) correspond to convex (concave) contour line toward the WNM.
%The gas diverges toward the WNM for the regions $K>0$ while
%the gas converges toward the CNM where $K<0$, 
%This behaviour is consistent with the velocity vectors in figure \ref{fig:denvel}a.

Note that the fast flows driven in the IFM are accompanied by a
fast phase transition between the CNM and IFM/WNM. 
If the kinetic energy of the fast flows in the IFM is carried into 
the CNM through the phase transition, it can act as a driving force of turbulence in the CNM.

%----------------------------------------------------------
\subsection{Mechanism of Driving and Dissipation of Kinetic Energy in Each Phase
}\label{sec:selfsus}
%----------------------------------------------------------
\begin{figure*}[htpb]
        \begin{center}
                \includegraphics[width=12cm]{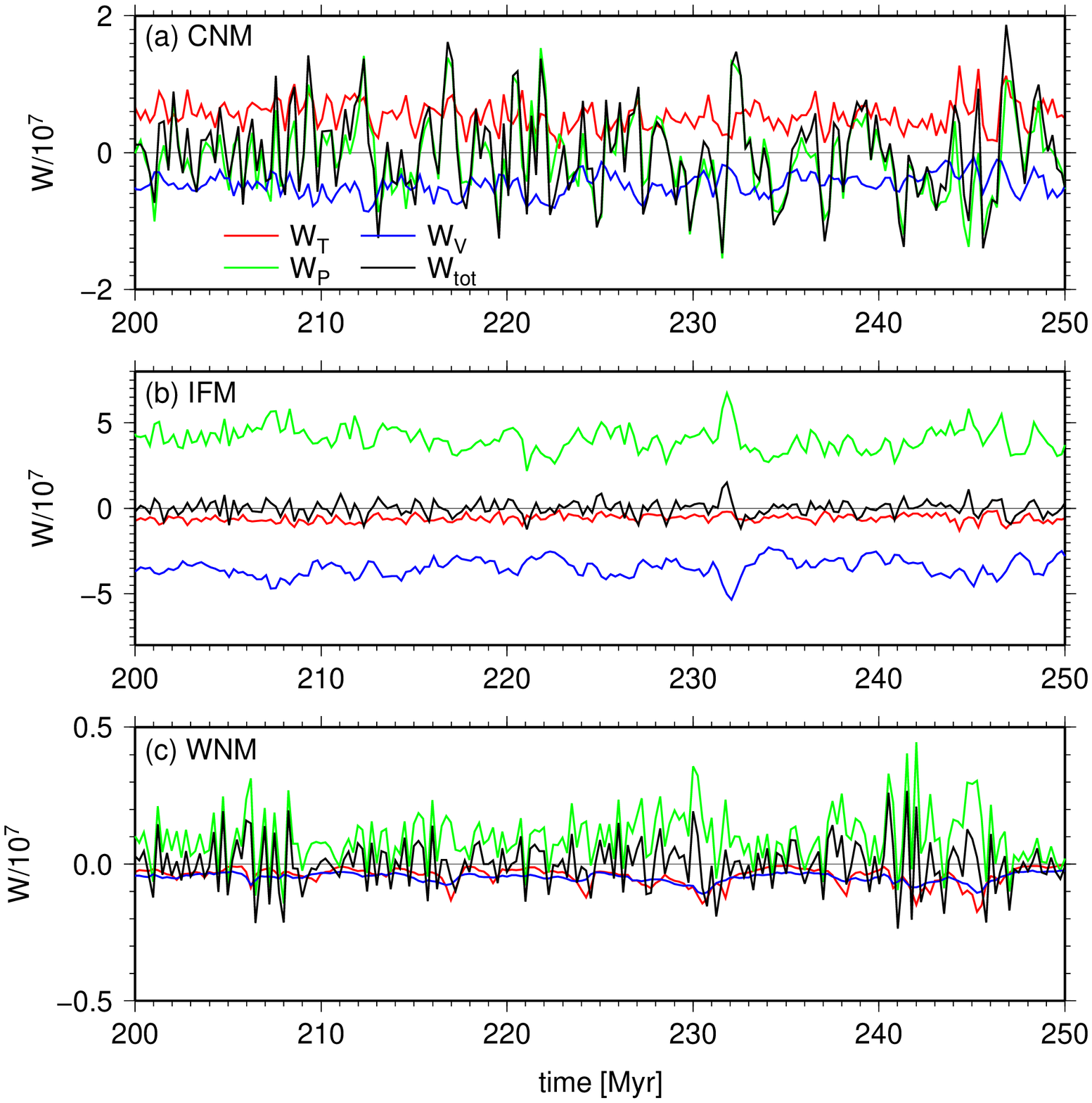}
        \end{center}
        \caption{
        Power contributions to the time evolution of the kinetic 
        energy for (a)the CNM, (b)IFM, and (c)WNM.
        In each panel, The red, green, and blue lines correspond to 
        $W_\mathrm{T}$, $W_\mathrm{P}$, and $W_\mathrm{V}$, 
        respectively, and the black line indicates 
        the total power, $W_\mathrm{tot}$. 
        The vertical coordinates are divided by $10^7$.
        }
        \label{fig:kincomp}
\end{figure*}

The turbulence seems to show saturation for $t>50\;\mathrm{Myr}$ (see Fig. \ref{fig:vel_las}).  
In the saturated state, driving of kinetic energy is expected to balance 
dissipation of kinetic energy in each phase.
In this section, we investigate what mechanism drives the kinetic energy and what mechanism dissipates it 
in each phase.
We consider 
the evolution equation for the total kinetic energy of phase $s=(\mathrm{cnm,\;ifm,\;wnm})$, given by 
\begin{equation}
        \frac{\partial }{\partial t} \int_{V_\mathrm{s}} \left( \frac{1}{2}\rho {\bf v}^2 \right)dV 
        =W_\mathrm{T,s} + W_\mathrm{P,s} + W_\mathrm{V,s} \equiv W_\mathrm{tot,s},
        \label{kinene4}
\end{equation}
where $W_\mathrm{P,s}$ and $W_\mathrm{V,s}$ indicate the powers due to the pressure gradient and 
viscous force, respectively, and are given by  
%%%%
\begin{equation}
        W_\mathrm{P,s} = - \int_{V_\mathrm{s}} dV {\bf v}\cdot {\bf \nabla}P,
\end{equation}
%%%%
and
%%%%
\begin{equation}
        W_\mathrm{V,s} = \int_{V_\mathrm{s}} dV v_\mu \nabla_{\nu} \sigma_{\mu\nu}.
\end{equation}
%%%%
Since the volume occupying the phase ``s'' changes with time, the interaction between 
adjoining phases should be taken into account.
This effect is represented by $W_\mathrm{T,s}$ and indicates the kinetic energy transport associated with
the phase transition. The detailed expression is given by 
\begin{equation}
     W_\mathrm{T,s} = - \oint_{\mathrm{int}} dS \frac{1}{2} \rho {\bf v}^2 \left( {\bf v}- {\bf v}_\mathrm{int} \right) \cdot{\bf n}_\mathrm{int},
\end{equation}
where $\oint_\mathrm{int} dS$ denotes the surface integral at the interface, 
${\bf v}_\mathrm{int}$ is the velocity at the interface, and
${\bf n}_\mathrm{int}$ is the normal unit vector at the interface pointing outwards with respect to 
the phase ``s''.
Since the kinetic energy is always positive, for $W_\mathrm{T,s}>0$ the kinetic energy injection from the other phases
to the given phase ``s'' exceeds the kinetic energy ejection from the phase ``s'' to the other phases,
and vice versa for $W_\mathrm{T,s}<0$.
The total power is denoted by $W_\mathrm{tot,s}$ in equation (\ref{kinene4}).  
The detailed derivation of equation (\ref{kinene4}) is described in Appendix \ref{sec:appkin}.

We numerically evaluate the powers in equation (\ref{kinene4}) at each time step.
There are two cautions. One is that numerical schemes that describe shock waves unavoidably 
contain numerical viscosity.
In Godunov's method adopted in this paper,  
the result from the nonlinear Riemann solver is used in the evaluation of the pressure gradient.
Thus, if a simple difference form is used for the evaluation of 
the pressure gradient in calculating $W_\mathrm{P,s}$, one misses the effects of numerical viscosity.
In this paper, the power due to numerical viscosity is calculated in the following way.
In calculating the pressure gradient at the cell center, the pressure at the cell boundary is required.
Godunov's method evaluates the pressure $P^*$ by using the result from the nonlinear Riemann solver, 
where the left- and right-hand side states are derived by an interpolation from cells.
The numerical viscous flux is evaluated approximately with $P^* - (P_\mathrm{L} + P_\mathrm{R})/2$, where 
$P_\mathrm{L}$ and $P_\mathrm{R}$ are the pressures in the left- and right-hand side states of the Riemann solver.
The powers due to numerical viscosity are evaluated from the numerical viscous flux.
The power $W_\mathrm{P,s}$ is calculated from $(P_\mathrm{L} + P_\mathrm{R})/2$.
Then, the power due to numerical viscosity is included in $W_\mathrm{V,s}$.
The other caution is that $W_\mathrm{T,s}$ is difficult to estimate directly using the finite-volume method. 
Thus, $W_\mathrm{T,s}$ 
is derived indirectly by subtracting $W_\mathrm{P,s}$ and $W_\mathrm{V,s}$ from the time derivative of the 
total kinetic energy of the phase ``s'' (see equation (\ref{kinene4})).

Fig. \ref{fig:kincomp} shows the time evolution of the three powers, 
$W_\mathrm{T}$, $W_\mathrm{P}$, and $W_\mathrm{V}$
for the CNM, IFM, and WNM for $200<t/\mathrm{Myr}<250$.
In each panel, the black line corresponds to the total power, $W_\mathrm{tot,s}$.
When $W_\mathrm{T,P,V}>0$ $(<0)$, the power increases (decreases) the kinetic energy.
To evaluate quantitatively which power is dominant in each of the driving and 
dissipation mechanisms, 
the time average of $W_\mathrm{T,P,V}$ is calculated in the temporal range $100<t/\mathrm{Myr}<530$.
The results are shown in Table \ref{table:1} for which the values have been normalized by 
$10^7m_\mathrm{H}L^2/\mathrm{Myr}$, where $L=2.4\mathrm{pc}$.
In each phase, the time average of the total powers, $\langle W_\mathrm{tot,s}
\rangle$, during $100<t/\mathrm{Myr}<530$ is almost zero, 
indicating that the turbulence reaches a quasi steady state. 

%%%%%%%%%%%%%%%%
\begin{table}
        \begin{center}
\begin{tabular}{|c|c|c|c|c|}
        \hline
        & $\langle W_\mathrm{T}\rangle$ & $\langle W_\mathrm{P}\rangle$ & $\langle W_\mathrm{V}\rangle$ 
        & $\langle W_\mathrm{tot}\rangle$ \\
        \hline
        CNM  & $1.1$ & $-0.045$ & $-1.1$ & $-0.009$\\
        IFM  & $-1.2$ & $9.0$ & $-7.8$ & $-0.00035 $\\
        WNM  & $-0.13$ & $0.27$ & $-0.14$& $-0.0005 $\\
        \hline
\end{tabular}
\caption{
Time average of powers for each phase. The values are normalized by 
$10^7 m_\mathrm{H}L^2/\mathrm{Myr}$, where $L=2.4\mathrm{pc}$.
}
\label{table:1}
        \end{center}
\end{table}
%%%%%%%%%%%%%%%%

Since the CNM provides the dominant contribution to the total kinetic energy of the system,
the driving mechanism of turbulence in the CNM is crucial.
Fig. \ref{fig:kincomp}a shows that $W_\mathrm{T,cnm}$ always takes a positive value. 
On the other hand, $W_\mathrm{P,cnm}$ fluctuates with a large amplitude around zero. 
This indicates that there are compressive waves inside the CNM.
As will be described later, the compressive waves are driven by the pressure decrement caused by conductive 
cooling in the IFM. 
Table \ref{table:1} shows that $\langle W_\mathrm{T,cnm} \rangle$ is balanced with viscous dissipation 
$\langle W_\mathrm{V,cnm}\rangle$ while
$\langle W_\mathrm{P,cnm}\rangle$ is almost zero.
Thus, it is confirmed that the main driver of turbulence in the CNM is 
kinetic energy injection from the IFM to CNM, as mentioned at the end of Section \ref{sec:selfsus}.

The kinetic energy in the CNM comes from the IFM through the phase transition.
Let us see the driving and dissipation mechanisms of the kinetic energy in the IFM.
Fig. \ref{fig:kincomp}b shows that only the pressure gradient force increases 
the kinetic energy (also see Table \ref{table:1}).
This pressure gradient arises from conductive cooling and heating near deformed 
CNM/IFM interfaces, and it accelerates the fluid.
On the other hand, kinetic energy is dissipated mostly by the viscosity when
$|\langle W_\mathrm{V,imf}\rangle|$ is slightly smaller than 
$\langle W_\mathrm{P,ifm}\rangle$ (see Table \ref{table:1}).
Since $W_\mathrm{T,ifm}$ is negative and $\langle W_\mathrm{T,ifm}\rangle\simeq-\langle W_\mathrm{T,cnm}\rangle$, 
the phase transition transports the small remaining kinetic energy from the IFM to the CNM.  
This kinetic energy injection drives turbulence in the CNM. 

Fig. \ref{fig:kincomp}c and Table \ref{table:1} show that 
the powers in the WNM are much smaller than those of the other two phases. 
As mentioned in Section \ref{sec:denvel}, the WNM is dissipative ($\mathrm{Re}<1$) for $L=2.4\;\mathrm{pc}$.
Thus, the WNM is not expected to contribute to the driving of turbulence.
The WNM is passively entrained by the fluid motion in the IFM.

\begin{figure*}[htpb]
        \begin{center}
                \includegraphics[width=14cm]{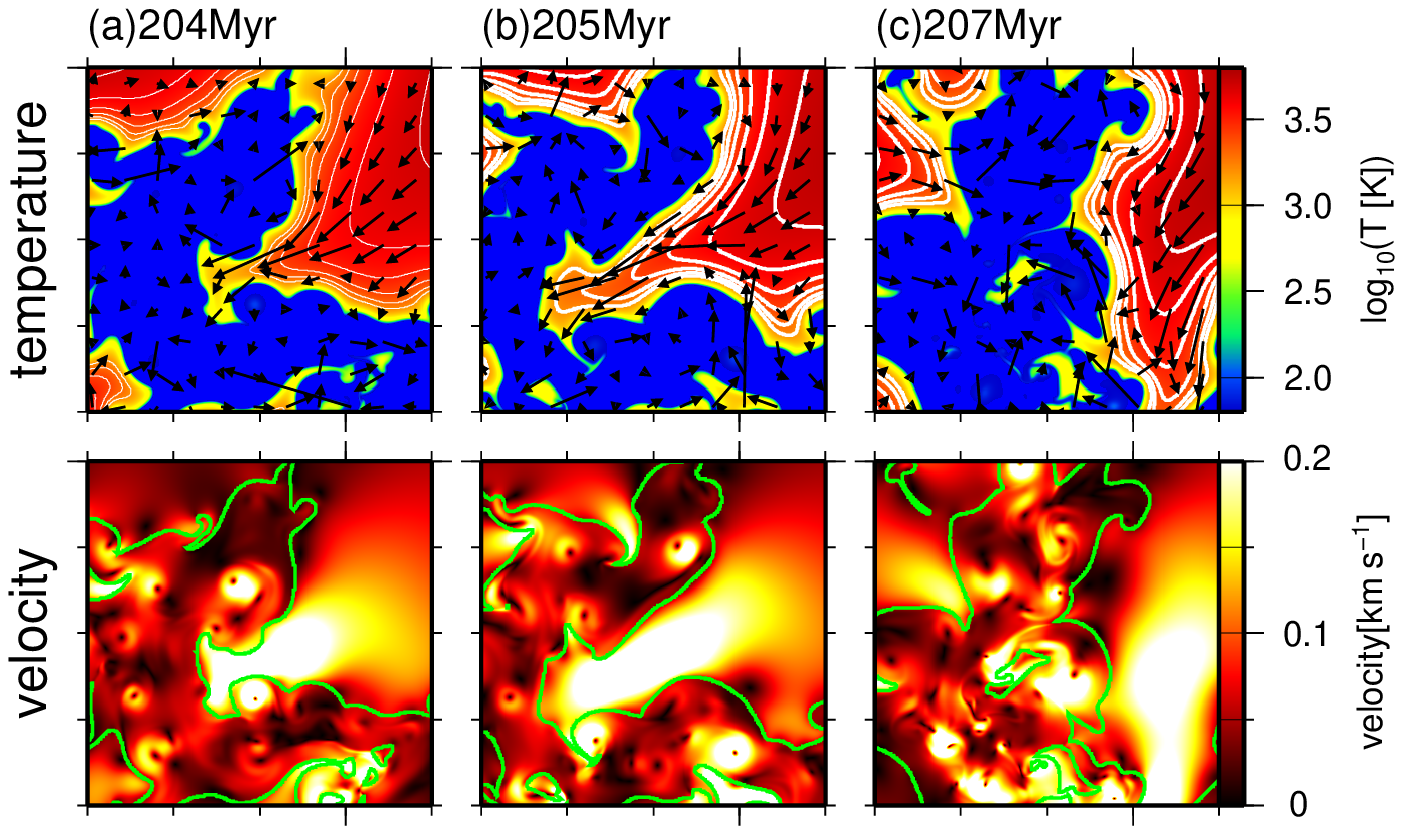}
        \end{center}
        \caption{
        Time sequences of temperature and velocity for $t=$(a)204Myr, (b)205Myr, and (c)207Myr
        in a rectangular box of size 0.4pc. 
        The arrows in the temperature map indicate the velocity field.
        The green lines in the velocity map denote the CNM/IFM interfaces.
        }
        \label{fig:seq}
\end{figure*}
%\begin{figure}[htpb]
%        \begin{center}
%                \includegraphics[width=8cm]{mech.eps}
%        \end{center}
%        \caption{
%        Schematic picture of kinetic energy injection mechanism from the IFM to CNM.
%        The dotted region corresponds to the CNM.
%        }
%        \label{fig:seq}
%\end{figure}

Fig. \ref{fig:kincomp} and Table \ref{table:1} suggest that turbulence in the CNM 
is mainly driven by kinetic energy injection through the phase transition from the IFM to the CNM.
To show this process more clearly, time sequences of the temperature  
and velocity are plotted for a rectangular region of size 0.4pc in Fig. \ref{fig:seq}.
The velocity distribution in Fig. \ref{fig:seq}a shows that there are many vortices inside the CNM 
which correspond to circular structures with central holes.
The turbulent motion in the CNM begins to pull its interface leftwards around the center in Fig. \ref{fig:seq}a.
Fig. \ref{fig:seq}b shows that the interface is largely stretched towards the CNM and 
a prominent concave CNM surface is formed.
As mentioned before, in the IFM enclosed by the concave CNM surface, thermal conduction cools the gas. 
This decreases the pressure in the IFM towards the CNM.
The resultant pressure gradient drives fast flows streaming into the concave CNM surface,
as shown in Fig. \ref{fig:kincomp}b.
%The flow continuously condenses onto the CNM.
In the concave CNM surface, one can see two parallel interfaces facing each other across the IFM, 
and one side of them is connected. 
The distance between them is about $0.1$ pc.
It is well known that two parallel interfaces approach and eventually annihilate \citep{ERS91} (also see Appendix \ref{app:anni}).
Fig. \ref{fig:seq}c shows the merging epoch.
During this merging process, the IFM between the two interfaces changes into the CNM, or the phase transition.
The conduction driven flow is almost parallel to the two interfaces. 
This means that the conduction driven flow is perpendicular to the approaching direction of the two interfaces.
Thus, this flow velocity in the IFM is almost preserved during the phase transition. 
Finally, the conduction driven flow in the IFM is incorporated into the kinetic energy of the CNM.

%%%%%
\begin{figure}[htpb]
        \begin{center}
                \includegraphics[width=7cm]{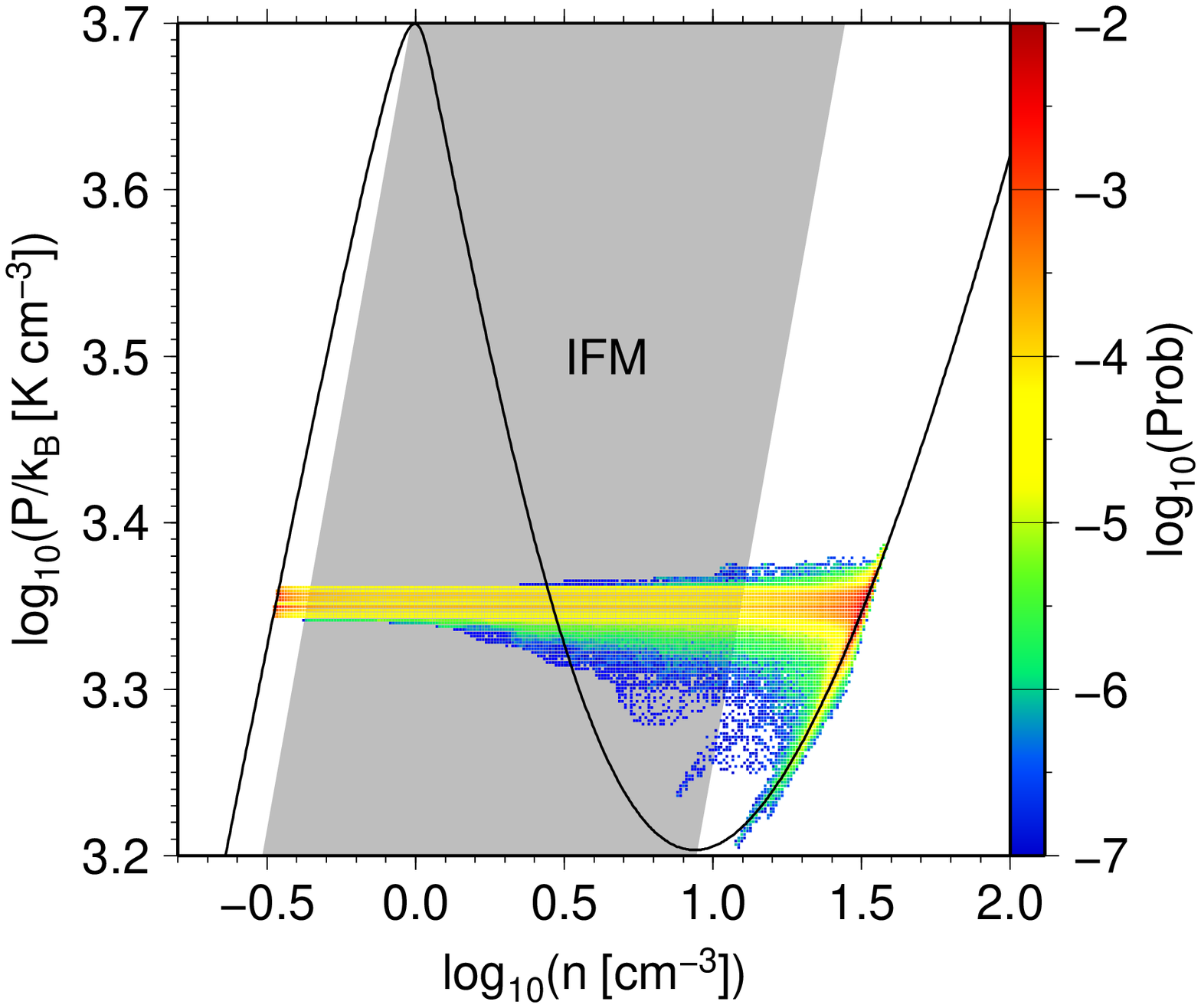}
        \end{center}
        \caption{
        Probability distribution in the $(n,P)$ plane averaged in the time 
        range of $200\le t/\mathrm{Myr}\le 220$.
        The gray region corresponds to the IFM.
        The unit is arbitrary. The black line corresponds to the thermal equilibrium curve.
        }
        \label{fig:pn}
\end{figure}
%%%%%
The variation of the pressure can be clearly seen in 
Fig. \ref{fig:pn}, which shows the probability distribution in the $(n,P)$ plane averaged over the time 
range $200\le t/\mathrm{Myr}\le 220$.
One can see that most of the gas resides at at constant pressure of $\log_{10}(P/k_\mathrm{B})\sim 3.35$.
However, low pressure regions are found in the IFM. 
This corresponds to the conductively cooled regions enclosed by 
concave CNM surfaces. Note that the CNM distributes along
the thermal equilibrium curve shown by the black line.
This is because the cooling/heating timescale of the CNM is so short that the CNM 
evolves along the thermal equilibrium curve. Moreover, the pressure distribution of the CNM extends downward 
in the $(n,P)$ plane.
As mentioned above, this large pressure variation in the CNM is attributed to 
conductive cooling in the IFM enclosed by the concave CNM surfaces.
This drives compressive waves in the CNM.

%--------------------------------------------
\section{Discussion}\label{sec:discuss}
%--------------------------------------------
%----------------------------------------------------------------------------------
\subsection{Self-Sustaining Mechanism of Turbulence}\label{sec:selfmech}
%----------------------------------------------------------------------------------
\begin{figure}[htpb]
        \begin{center}
                \includegraphics[width=7cm]{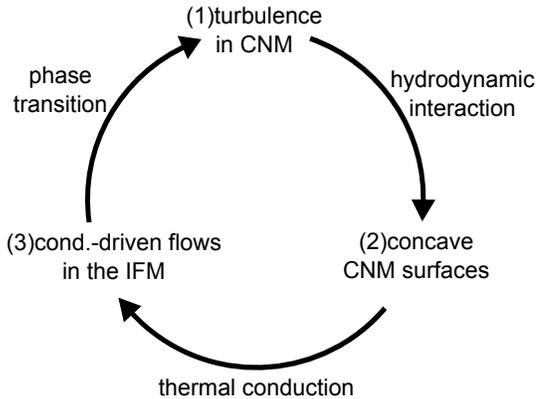}
        \end{center}
        \caption{
        Schematic picture of self-sustaining mechanism. 
        }
        \label{fig:mech}
\end{figure}

From the findings in Section \ref{sec:result}, the following self-sustaining mechanism 
is realized in bistable turbulence.
The self-sustaining mechanism can be divided into three parts, as shown in Fig. \ref{fig:mech}.
(1)Turbulence inside the CNM deforms its CNM surface and (2)creates concave CNM surfaces.
In the IFM enclosed by the concave CNM surfaces, 
(3)thermal conduction drives flows that stream towards the concave CNM surfaces.
The kinetic energy of the IFM flows is transported into the CNM through the phase transition.
In this manner, the cycle in Fig. \ref{fig:mech} is self-sustained in a bistable system.

%----------------------------------------------------------------------------------
\subsection{Typical Time and Length Scales of Kinetic Energy Injection}\label{sec:typ_time}
%----------------------------------------------------------------------------------
\begin{figure}[htpb]
        \begin{center}
                \includegraphics[width=7cm]{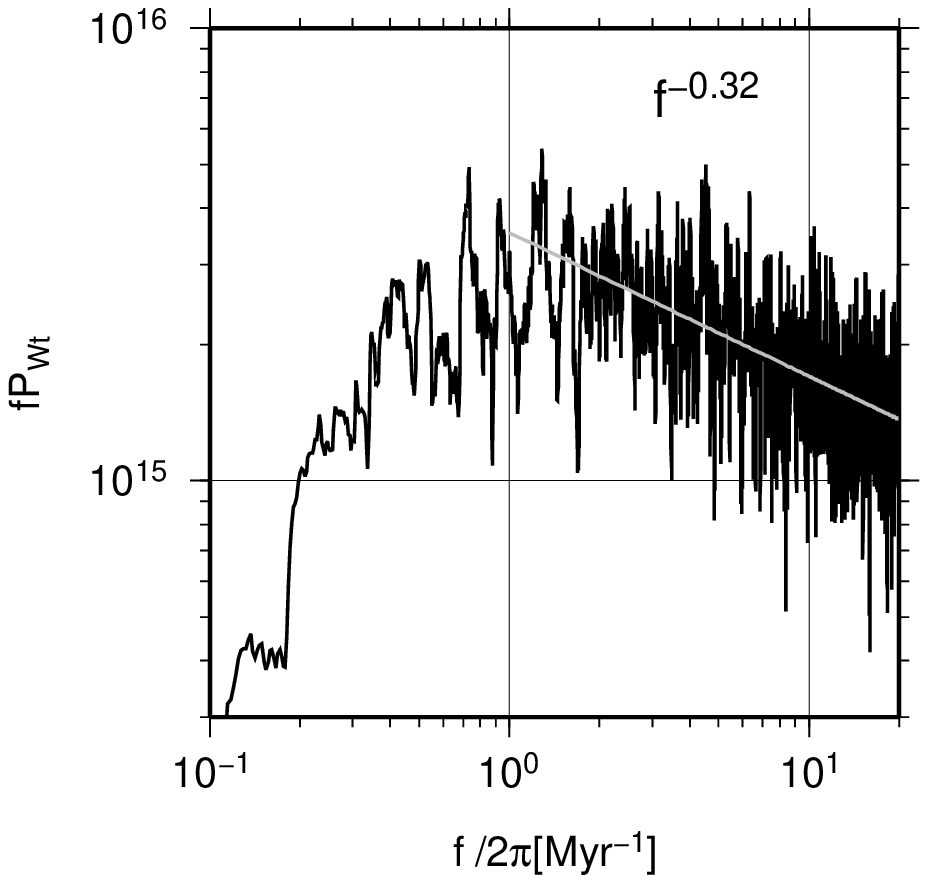}
        \end{center}
        \caption{
        Power spectrum $f\times P_{W_\mathrm{T,cnm}}(f)$ of $W_\mathrm{T,cnm}$.
        The gray line corresponds to a fitting formula for $f/2\pi> 2$ that is proportional 
        to $f^{-0.32}$.
        }
        \label{fig:spec}
\end{figure}

From Section \ref{sec:selfsus}, it is found that kinetic energy injection from the IFM to the CNM 
occurs in concave CNM surfaces. 
To investigate the typical timescale of kinetic energy injection, the power spectrum of $W_\mathrm{T,cnm}$ is 
plotted in Fig. \ref{fig:spec}. The vertical axis denotes the power spectrum multiplied by the frequency.
From Fig. \ref{fig:spec}, $f\times P_{W_\mathrm{T,cnm}}(f)$ peaks around $f/2\pi \sim 1\;\mathrm{Myr^{-1}}$, 
and has a power law $\propto f^{-0.32}$ in the high frequency limit. 
Thus, kinetic energy injection with a timescale of $2\pi/f \sim 1\;\mathrm{Myr}$ provides a dominant 
contribution to $W_\mathrm{T,cnm}$.

The timescale of energy injection is expected to be comparable to that of annihilation of 
two parallel interfaces, given by $t_\mathrm{merge}=1.47(e^{d/0.14\;\mathrm{pc}}-1)\;\mathrm{Myr}$, where 
$d$ is the distance between two interfaces (see Appendix \ref{app:anni}).
From this fitting equation, $d$ is about 0.1 pc for $t_\mathrm{merge}=1\;\mathrm{Myr}$. 
Assuming that the timescale of kinetic energy injection is determined by the annihilation of two parallel 
interfaces, the typical scale of the concave CNM surfaces is expected to be 0.1 pc.
From Figs. \ref{fig:denvel}, \ref{fig:denzoom}, and \ref{fig:seq}, the typical scale of the prominent 
concave CNM surfaces appears to be consistent with $\sim 0.1$ pc.
Interestingly, the typical scale of $0.1$ pc is comparable to the thickness of the IFM in a plane-parallel geometry
(see Fig. \ref{fig:eq}).

%----------------------------------------------------------------------------------
\subsection{Dependence of Saturation Level on Simulation Box Size}\label{sec:Ldep}
%----------------------------------------------------------------------------------
\begin{figure}[htpb]
        \begin{center}
                \includegraphics[width=7cm]{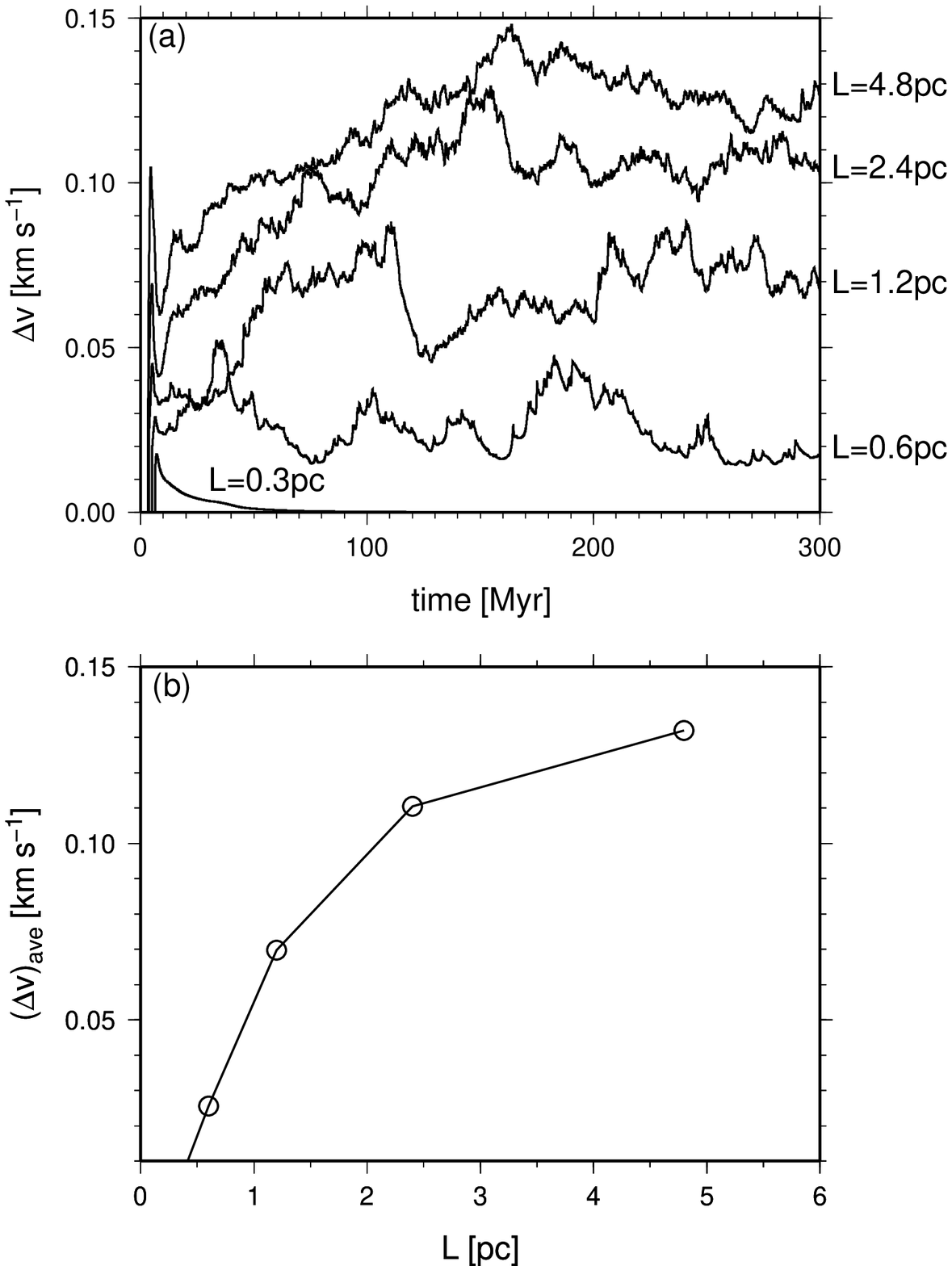}
        \end{center}
        \caption{
        (a)Time evolution of velocity dispersion. 
        From bottom to top, the lines correspond to $L/\mathrm{pc}=0.3$, 0.6, 1.2, 2.4, and 4.8.
        (b)Box-size dependence of the time averaged velocity dispersion $(\Delta v)_\mathrm{ave}$.
        }
        \label{fig:Ldep}
\end{figure}

In this section, the box size dependence of turbulence is considered.
The simulations are performed under the same initial condition while keeping 
the grid size constant for various box sizes.
Fig. \ref{fig:Ldep}a shows the time evolution of the velocity dispersion for $L=$0.3, 0.6, 1.2, 
2.4, and 4.8 pc. 
Only for the smallest box size ($L=0.3$ pc) does the turbulence decay.
For larger box sizes, the velocity dispersion initially increases and 
saturates around $t\sim 100\;\mathrm{Myr}$. 
Figure \ref{fig:Ldep}a reveals that the saturation levels increase with $L$.
This trend can be understood from the driving mechanism of turbulence.
If the simulation box is too small, the effect of conductive cooling can traverse
the simulation box size within the survival time of curved interfaces, and 
a quasi-isobaric state is quickly established. 
As a result, driving of fast flows is limited, leading to 
a small turbulent velocity. 

Fig. \ref{fig:Ldep}b shows the box size dependence of the saturation level.
The vertical axis indicates the time averaged velocity dispersion for $100<t/\mathrm{Myr}<300$.
The averaged velocity dispersion seems to saturate for $L\ge 4.8\;\mathrm{pc}$.
This saturation can be explained from 
the typical timescale of the energy injection ($\sim 1\;\mathrm{Myr}$) found in Section \ref{sec:typ_time}. 
The sound crossing length within $1\;\mathrm{Myr}$ is as large as $\sim 7$ pc, where the sound speed of the WNM 
($\sim 7\;\mathrm{km\;s^{-1}}$) is used.
Thus, for $L>7\;\mathrm{pc}$, the velocity dispersion is expected to be independent of the box size.
This is consistent with the fact that $(\Delta v)_\mathrm{ave}$ begins to saturate around $L=2.4\sim 4.8$ pc.
The saturation level is expected to be $0.14\sim 0.15$ km s$^{-1}$ by extrapolation of Fig. \ref{fig:Ldep}b.
Furthermore, from the driving mechanism, the velocity dispersion is not expected to be larger than the conduction driven flow.
The expected saturation level is consistent with the typical velocity shown in equation (\ref{curve}). 

%---------------------------------------------------------
\subsection{Comparison with Previous Works}
%---------------------------------------------------------
\citet{KI06} have investigated the dependence of the saturation level on the simulation box, and 
found that the velocity dispersion saturates for $L>35\;\mathrm{pc}$, and 
the saturation level is $\sim 0.3$ km s$^{-1}$. 
This conclusion appears to contradict our result in Section \ref{sec:Ldep}.
This is attributed to the fact that they adopted an artificially large thermal conductivity and viscosity 
for calculations with larger boxes \citep{KI06}.
We have done the same calculation for the case with $L=144$pc.
After 100 Myr, when they terminate the calculation, we find that the turbulence decays because of artificially large viscosity.
The saturation found by \citet{KI06} may come from the initial growth of the TI.

\citet{Brandenburg07} performed similar calculations with a different analytic net cooling rate.
As mentioned in Section \ref{sec:intro}, they used an artificially large thermal conductivity and 
viscosity that are proportional to density. 
If we use the same conditions in our two-dimensional calculations, we also observe decaying turbulence. 
To drive turbulence, it is important that turbulence in the CNM deforms the 
interface. However, because of the large viscosity, the turbulence quickly decays.
Another difference between our work and theirs is the dimenstionality.
This will be discussed in the next section.

%---------------------------------------------------------
\subsection{Three-dimensional Case Without Magnetic Field}
%---------------------------------------------------------
In this paper, we focus on the two-dimensional evolution of a
bistable gas. It is well known that the evolution of vorticity
strongly depends on the dimensionality.
In two dimensions, enstrophy is conserved.
Thus, the vortex filaments behave like particles. 
On the other hand, in three dimensions, a vortex cascades into smaller 
vortices though stretching.
Note that the deformation of CNM surfaces requires relatively large vortices in the CNM.
For three dimensions, 
it is expected that the deformation of CNM surfaces becomes inefficient.
Thus, it is possible that the driving mechanism proposed in this paper cannot 
maintain turbulence in three dimensions.
However, no one has performed three-dimensional simulations with sufficient resolution to resolve the thickness 
of the CNM/IFM interface \citep{KI04} because the required computational cost is enormous.
To simulate the three dimensional simulation efficiently,  
one of the promising methods is adaptive mesh refinement technique, for example, 
with a refinement criterion based on the local Field length.
The effect of three dimensions is beyond the scope of this paper, but should 
be investigated in forthcoming work.

%---------------------------------------------------------
\subsection{Effect of Magnetic Field}
%---------------------------------------------------------
In this paper, for simplicity
we focus on the two-dimensional hydrodynamical evolution of 
a bistable fluid without magnetic fields.
However, in realistic situations, magnetic fields play important roles in the dynamics 
of a bistable fluid because the typical magnetic field strength of HI gas is 
about a few micro Gauss \citep{HT05}.
The one-dimensional evolution of magnetized bistable gas has been investigated by 
\citet{IIK07} and \citet{SZ10}. They show that ambipolar diffusion efficiently transports
magnetic field across a transition layer, leading to a flat magnetic field strength profile.
However, the multi-dimensional evolution is still unclear.
The plasma beta of HI gas is less than or comparable to unity.
Therefore, the turbulent velocity is less than the Alfv{\'e}n speed of
$\sim 1\;\mathrm{km\;s^{-1}}$ for the CNM and $\sim 10\;\mathrm{km\;s^{-1}}$ for the WNM.
This corresponds to weak Alfv{\'e}nic turbulence where the energy preferentially cascades 
in directions perpendicular to the mean magnetic field in the ideal MHD limit
\citep{SG94}. Thus, the outcome can be similar to two-dimensional 
hydrodynamic turbulence. 
The effects of magnetic fields will be investigated in a forthcoming paper.

%----------------------------------------------------------------------------------
\subsection{Implications for Interstellar Turbulence}\label{sec:implication}
%----------------------------------------------------------------------------------
The self-sustaining mechanism drives turbulence in the CNM at the level of about $0.1-0.2\;\mathrm{km\;s^{-1}}$.
It is well known that the velocity dispersion 
($\sim1\;\mathrm{km\;s^{-1}}$ at $1\;\mathrm{pc}$) in the real ISM is 
much larger than that found in this paper \citep[e.g.,][]{L81,HF12}. 
Thus, the turbulence analyzed in the paper cannot alone explain the interstellar turbulence quantitatively.

The difference between turbulence in this paper and the interstellar turbulence 
could be a consequence of the oversimplified setup in the paper.
In turbulence under periodic boundary condition without any dynamical forcing,
the two stable phases (WNM/CNM) are completely separated and the unstable gas resides only in the interfaces between 
the WNM/CNM. In this case, as shown in Fig. \ref{fig:ent}, 
thermal conduction is the most important thermal process,
it drives only weak turbulence with $\Delta v= 0.1-0.2\;\mathrm{km\;s^{-1}}$.

In realistic astrophysical environments, on the other hand, the ISM is frequently disturbed and compressed
by energetic phenomena the length scales of which are larger than a few parsec, 
such as expansions of HII regions and supernova explosions with timescale on 
the order of 1Myr \citep{MO77} that is 
comparable to the typical time scale of the kinetic energy injection derived in Section \ref{sec:typ_time}. 
Thus, the large scale disturbances should be important for the interstellar turbulence.
\citet{KI02} have demonstrated that the shock compression of the ISM induces turbulence which is composed of
shocked warm gas and cold cloudlets. The velocity dispersion is comparable to the observed values.
Recently, similar calculations have been performed by many authors \citep{AH05,HA07,Hetal05,Hetal06,Vetal06,Vetal07,II08,II09,
Hetal09,Betal09,Vetal11,InoueI12}.
However, the detailed mechanism to create turbulent structure is not fully analysed because of its complexity.

One important difference between turbulence in shocked ISM and this paper is 
the physical properties of the unstable gas (IFM). 
The shock heated gases are thermally unstable and subject to strong radiative cooling by which 
the CNM is formed.
This cooling drives gas flows that reach a velocity as large as $\sim$ several km s$^{-1}$.
Since this typical velocity is large, the Reynolds number becomes high enough for turbulence to be driven.
Unlike the turbulence in this paper, thermal conduction is less important because of the strong radiative cooling.
In addition to the TI (strong radiative cooling),
turbulence in the unstable gas is driven by vortex generation due to the baroclinic effect, 
the Richtmyer-Meshkov instability \citep{IYI09,IYI12,InoueI12,Setal12}. 

The results in Sections \ref{sec:selfsus} show that the kinetic energy transfer from the unstable gas 
to the CNM through the phase transition is the main driving source of turbulence in the CNM.
The same process can be expected in realistic situations with strong shocks,
in addition to other possible driving mechanisms of turbulence, such as the Kelvin-Helmholtz and Rayleigh-Taylor instabilities.
Since the velocity dispersion of the unstable gas is larger ($\sim$ several km s$^{-1}$), the transferred 
turbulent kinetic energy is expected to be larger. The resultant velocity dispersion inside individual CNM clouds 
may be supersonic (sound speed of CNM $\sim$ 0.2 km s$^{-1}$). 
This mechanism may also affect molecule formation inside the CNM because of the mixing between 
fresh CNM and preexisting CNM. These subjects will be discussed in a forthcoming paper.

%--------------------------------------------
\section{Summary}\label{sec:summary}
%--------------------------------------------
In this paper, we have investigated the turbulent structure of the bistable ISM by
using two-dimensional hydrodynamic simulations with a realistic cooling rate, thermal conduction, 
and physical viscosity. 
Our results are summarized as follows:

\begin{enumerate}
        \item It is confirmed that turbulence is sustained for at least 500 Myr without any dynamical forcing. 
              The velocity dispersion of the IFM is comparable to that of the CNM while
              that of the WNM is only half of that of the other two phases.
              The dominant contribution to the velocity dispersion is provided by the CNM because of 
              its large mass fraction. 

        \item Fast flows are observed in the IFM near strongly deformed CNM/IFM interfaces.
              There are two prominent flows. First, gas in the IFM streams into 
              concave CNM surfaces.
              Second, gas in the IFM flows towards the WNM from pillars of the CNM.
              It is found that these fast IFM flows are driven by thermal conduction.

        \item The mechanisms of driving and dissipation of kinetic energy are investigated in the saturation state
              of the three phases.
              In the CNM, the dominant driving mechanism is kinetic 
              energy injection from the IFM through the phase transition.
              This injected kinetic energy comes from the fast flows driven by strong conductive cooling near 
              concave CNM surfaces. 
              Since the IFM near concave CNM surfaces is surrounded by the CNM,
              the pressure drop due to conductive cooling in the IFM induces relatively 
              large pressure fluctuations in the CNM.

      \item A self-sustaining mechanism of bistable turbulence is summarized in Fig. \ref{fig:mech}.
            Turbulence inside the CNM creates concave CNM surfaces.
            Fast flows driven by thermal conduction in the IFM stream into the concave CNM surfaces 
            and their kinetic energy is transported into 
            the CNM through the phase transition. 
            In this way, the deformation of CNM surfaces by turbulence eventually enhances its kinetic energy.
            The free energy of this driving mechanism originally comes from the external heating that maintains 
            the temperature difference between the CNM/WNM. This temperature difference drives flows in the IFM that 
            become the driving source of turbulence in the CNM. 
            
\end{enumerate}

%---------------------------------------
\section*{Acknowledgments}
%---------------------------------------
We thank the anonymous referee for many constructive comments that improve this paper significantly.
We thank Dr. Tsuyoshi Inoue and Dr. S. Toh for valuable discussions.
We also thank Dr. Jannifer M. Stone for valuable discussions and careful reading this manuscript.
Numerical computations were carried out on Cray XT4 and XC30 at the CfCA of the National Astronomical 
Observatory of Japan and SR16000 at YITP in Kyoto University.
KI is supported by a Research Fellowship from the Japan Society for the Promotion of Science
for Young Scientists.
SI is supported by Grants-in-Aid for Scientific Research from the MEXT of Japan (23244027 and 23103005). 

\appendix
%--------------------------------------------------------------
\section{Derivation of Evolution Equation for Total Kinetic Energy of Three Phases}\label{sec:appkin}
%--------------------------------------------------------------
From equation (\ref{eoc}) and (\ref{eom}), the evolution equation 
for the kinetic energy is given by 
\begin{equation}
        \frac{\partial }{\partial t} \left( \frac{1}{2}\rho {\bf v}^2 \right)
         +{\bf \nabla} \cdot \left( \frac{1}{2}\rho {\bf v}^2 {\bf v}\right)
         =- {\bf v}\cdot {\bf \nabla}P + v_\mu \nabla_{\nu} \sigma_{\mu\nu}.
         \label{kinevo}
\end{equation}
In this appendix, the evolution equation for the total kinetic energy is derived for each of the three phases.
The procedure of the derivation is the same for all three phases.
Thus, the phase ``s'' is considered, where ``s'' denotes the label of the phase (CNM, IFM, and WNM).

The whole domain is divided into two subdomains: the phase ``s'' and the other phases.
We introduce a scalar field $\psi_\mathrm{s}$ given by 
\begin{equation}
        \psi_\mathrm{s}(t,{\bf x}) = \left\{
        \begin{array}{cl}
                1 & \mbox{inside the phase ``s''} \\
                0 & \mbox{elsewhere}
        \end{array} 
        \right..
\end{equation}
Using $\psi_\mathrm{s}$, one can define a normal unit vector at the interface pointing outwards with respect to 
the phase ``s'',
\begin{equation}
        {\bf n}_\mathrm{int} = - \left(\frac{\nabla \psi_\mathrm{s}}{|\nabla \psi_\mathrm{s}|}\right)_\mathrm{int},
        \label{nb}
\end{equation}
where the subscript ``int'' denotes the value at the interface.
For an observer moving with the interface, $\psi_\mathrm{s}$ does not change in time. Thus,
$\psi_\mathrm{s}$ obeys 
\begin{equation}
        \frac{\partial \psi_\mathrm{s}}{\partial t} + {\bf v}_\mathrm{int}\cdot {\bf \nabla} \psi_\mathrm{s} = 0,
        \label{psi}
\end{equation}
where ${\bf v}_\mathrm{int}$ is the velocity of the interface and is zero everywhere except at
the interface.

The time evolution of the total kinetic energy of the phase ``s'' is 
given by 
\begin{equation}
        \frac{\partial }{\partial t} \int dV \psi_\mathrm{s} E_\mathrm{kin}
        = 
        \int dV \psi_\mathrm{s} \frac{\partial  E_\mathrm{kin}}{\partial t}+ 
        \int dV \frac{\partial \psi_\mathrm{s}}{\partial t} E_\mathrm{kin}, 
        \label{kinene}
\end{equation}
where $E_\mathrm{kin}=\rho {\bf v}^2/2$.
The first term on the right-hand side of equation (\ref{kinene}) is considered.
Using equation (\ref{kinevo}), one gets
\begin{eqnarray}
        \int dV \psi_\mathrm{s} \frac{\partial E_\mathrm{kin}}{\partial t}
    &=&
    -\int dV \psi_\mathrm{s} {\bf \nabla} \cdot \left( E_\mathrm{kin} {\bf v}\right)\nonumber \\
    &-& \int dV \psi_\mathrm{s} {\bf v}\cdot {\bf \nabla}P \nonumber \\
    &+ &\int dV \psi_\mathrm{s} v_\mu \nabla_{\nu} \sigma_{\mu\nu}.
        \label{kinene1}
\end{eqnarray}
The integrand of the first term on the right-hand side of equation (\ref{kinene1}) can be rewritten as 
\begin{eqnarray}
        \psi_\mathrm{s} {\bf \nabla} \cdot \left( E_\mathrm{kin} {\bf v}\right)
        &=&  {\bf \nabla} \cdot \left( \psi_\mathrm{s} E_\mathrm{kin} {\bf v}\right) - E_\mathrm{kin} {\bf v}\cdot {\bf \nabla} \psi_\mathrm{s}    
    \nonumber\\
    &=&  {\bf \nabla} \cdot \left( \psi_\mathrm{s} E_\mathrm{kin} {\bf v}\right) + E_\mathrm{kin} {\bf v}\cdot 
    n_\mathrm{int}|{\bf \nabla} \psi_\mathrm{s}|,
     \label{eq1}
\end{eqnarray}
where equation (\ref{nb}) is used in the last line.
From Gauss's theorem, the volume integral of the first term on the right-hand side of equation (\ref{eq1})
vanishes because of the periodic boundary conditions.
From equations (\ref{kinene1}), (\ref{eq1}), and (\ref{psi}), equation (\ref{kinene}) becomes
\begin{eqnarray}
        \frac{\partial }{\partial t} \int dV \psi_\mathrm{s} E_\mathrm{kin}
        &=&- \int dV E_\mathrm{kin} \left( {\bf v}- {\bf v}_\mathrm{int} \right)\cdot{\bf n}_\mathrm{int} |{\bf \nabla}\psi_\mathrm{s}|\nonumber \\
        &-& \int dV \psi_\mathrm{s} {\bf v}\cdot {\bf \nabla}P + \int dV \psi_\mathrm{s} v_\mu \nabla_{\nu} \sigma_{\mu\nu}.
        \label{kinene2}
\end{eqnarray}
Since $|{\bf \nabla}\psi_\mathrm{s}|$ is a delta function that is infinity at the interface and zero elsewhere,
equation (\ref{kinene1}) becomes 
\begin{equation}
        \frac{\partial }{\partial t} \int dV \psi_\mathrm{s} E_\mathrm{kin}
        =W_\mathrm{T,s} + W_\mathrm{P,s} + W_\mathrm{V,s}
        \label{kinene3}
\end{equation}
where
\begin{equation}
  W_\mathrm{T,s} = - \oint_\mathrm{int} dS E_\mathrm{kin} \left( {\bf v}- {\bf v}_\mathrm{int} \right) \cdot{\bf n}_\mathrm{int},
\end{equation}
\begin{equation}
        W_\mathrm{P,s} = - \int dV \psi_\mathrm{s} {\bf v}\cdot {\bf \nabla}P,
\end{equation}
\begin{equation}
        W_\mathrm{V,s} = \int dV \psi_\mathrm{s} v_\mu \nabla_{\nu} \sigma_{\mu\nu},
\end{equation}
and $\oint_\mathrm{int} dS$ denotes the surface integral at the interface, 
$W_\mathrm{T,s}$ represents the 
kinetic energy transport across the interface relating to the phase transition. 
$W_\mathrm{P,s}$ and $W_\mathrm{V,s}$ correspond to powers due to the pressure gradient and 
viscous force, respectively.
The integral $\int dV \psi_s$ is the same as $\int_{V_\mathrm{s}} dV$ used in Section \ref{sec:selfsus}.

%------------------------------------------------------------------------------------------------------
\section{Annihilation of Two Parallel Interfaces in Plane-Parallel Geometry}\label{app:anni}
\begin{figure}[htpb]
        \begin{center}
                \includegraphics[width=7cm]{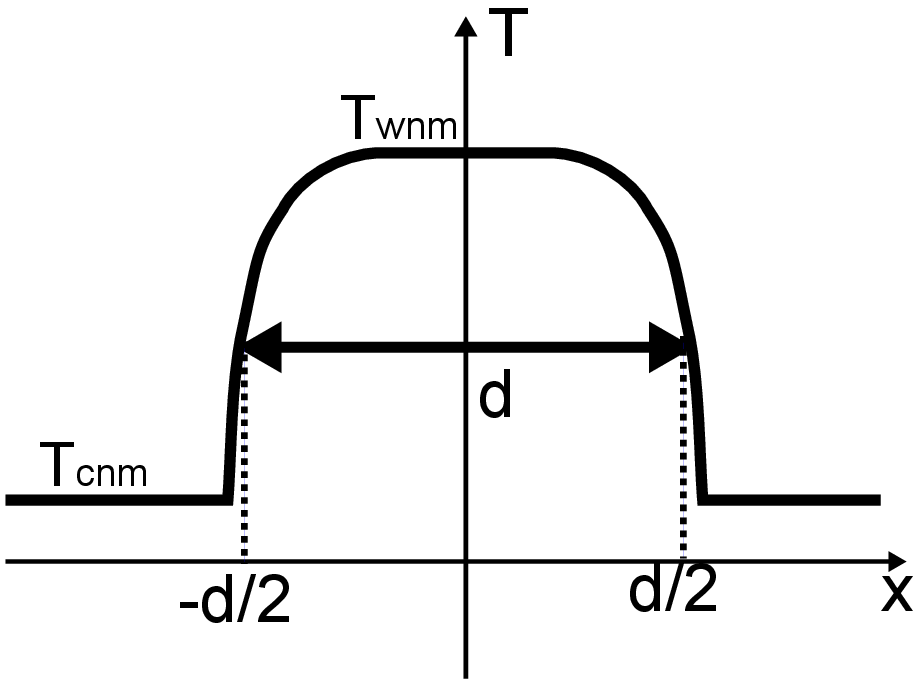}
        \end{center}
        \caption{
        Schematic picture of initial condition.
        }
        \label{fig:inianni}
\end{figure}

%------------------------------------------------------------------------------------------------------
In this Appendix, to derive the merging timescale of two parallel interfaces as a function of 
the distance between them,
a one-dimensional simulation is performed.
The temperature distribution of the static solution at $P=P_\mathrm{sat}$ \citep{ZP69} is denoted by 
$T_\mathrm{ZP}(x)$, that is $T_\mathrm{cnm}$ for $x\rightarrow-\infty$ and 
$T_\mathrm{wnm}$ for $x\rightarrow \infty$.
The origin of $T_\mathrm{ZP}$ is defined at the point where $T_\mathrm{ZP}(x=0)=(T_\mathrm{cnm}+T_\mathrm{wnm})/2$. 
An initial temperature distribution is assumed as follows:
\begin{equation}
        T(x) = T_\mathrm{ZP}(x+d/2) + T_\mathrm{ZP}(-x+d/2) - T_\mathrm{wnm},
\end{equation}
where $d$ is the distance between two interfaces.
Fig. \ref{fig:inianni} shows a schematic picture of the initial condition.
The WNM is sandwiched by two CNM phases.
The velocity is assumed to be zero and the pressure is constant.

\begin{figure}[htpb]
        \begin{center}
                \includegraphics[width=7cm]{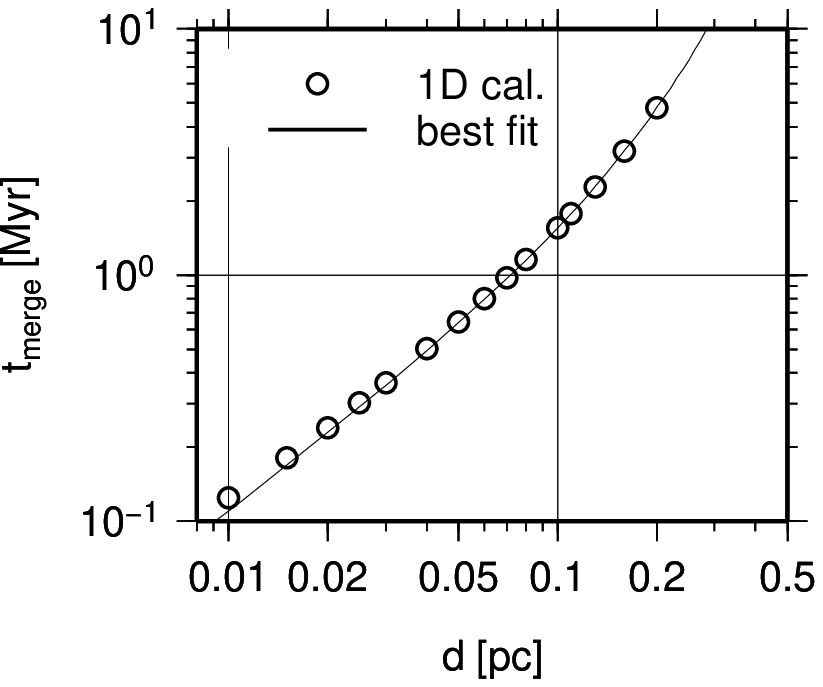}
        \end{center}
        \caption{
        Merging timescale of two parallel interfaces.
        The circles indicate the results of the 1D calculations and 
        the line corresponds to the fitting formula.
        }
        \label{fig:anni}
\end{figure}

The interfaces approach each other and eventually merge.
The merging time is estimated using the 1D simulations for various $d$ and is plotted in 
Fig. \ref{fig:inianni}.
It is seen that the merging time increases with $d$.
From perturbation theory, \citet{ERS91} derived an analytic formula 
($t_\mathrm{merge}=t_0\left( \exp(d/\lambda_0) -1 \right)$),
where $\lambda_0$ is the thickness of the transition layer and $t_0$ is a typical timescale.
The simulation points in Fig. \ref{fig:anni} are fitted by this analytic formula quite well. 
It is found that $t_0=1.47\;\mathrm{Myr}$ and $\lambda_0=0.14\;\mathrm{pc}$ which is consistent with 
the thickness of the IFM (0.1 pc).

%------------------------------------------------------------------
\bibliographystyle{apj}

\end{document}